\documentclass[]{pasj01}

\Received{2019/11/22}
\Accepted{2020/01/29}
 
\begin{document} 

\title{Deeply cooled core of the Phoenix galaxy cluster imaged by ALMA
with the Sunyaev-Zel'dovich effect}

\author{Tetsu \textsc{Kitayama}\altaffilmark{1},
 Shutaro \textsc{Ueda}\altaffilmark{2},
 Takuya \textsc{Akahori}\altaffilmark{3},
 Eiichiro \textsc{Komatsu}\altaffilmark{4,5},   
 Ryohei \textsc{Kawabe}\altaffilmark{6,7,8},
 Kotaro \textsc{Kohno}\altaffilmark{9,10},
 Shigehisa \textsc{Takakuwa}\altaffilmark{11},
 Motokazu \textsc{Takizawa}\altaffilmark{12},
 Takahiro \textsc{Tsutsumi}\altaffilmark{13},
 and Kohji \textsc{Yoshikawa}\altaffilmark{14}
}
\altaffiltext{1}{Department of
  Physics, Toho University, Funabashi, Chiba 274-8510, Japan}
\altaffiltext{2}{Academia Sinica Institute of Astronomy and Astrophysics
 (ASIAA), No. 1, Section 4, Roosevelt Road, Taipei 10617, Taiwan}
\altaffiltext{3}{Mizusawa VLBI observatory, National Astronomical
 Observatory of Japan, 2-21-1 Osawa, Mitaka, Tokyo 181-8588, Japan} 
\altaffiltext{4}{Max-Planck-Institut f\"{u}r Astrophysik, Karl-Schwarzschild Str. 1, D-85741 Garching, Germany}
\altaffiltext{5}{Kavli Institute for the Physics and Mathematics of the Universe (Kavli IPMU, WPI), Todai Institutes for Advanced Study, the University of Tokyo, 5-1-5 Kashiwanoha, Chiba 277-8583, Japan} 
 \altaffiltext{6}{National Astronomical Observatory of Japan, 2-21-1 Osawa,
Mitaka, Tokyo 181-8588, Japan}
\altaffiltext{7}{The Graduate University for Advanced Studies (SOKENDAI), 
2-21-1 Osawa, Mitaka, Tokyo 181-8588, Japan}
\altaffiltext{8}{Department of Astronomy, 
The University of Tokyo, 7-3-1 Hongo, Bunkyo, Tokyo 113-0033, Japan}
\altaffiltext{9}{Institute of Astronomy, School of Science, The University of Tokyo, 2-21-1 Osawa, Mitaka, Tokyo 181-
0015, Japan}
\altaffiltext{10}{Research Center for the Early Universe, School of Science, 
The University of Tokyo, 7-3-1 Hongo, Bunkyo, Tokyo 113-0033, Japan}
\altaffiltext{11}{Department of Physics and Astronomy, Graduate School of Science and Engineering,
Kagoshima University, 1-21-35 Korimoto, Kagoshima, Kagoshima 890-0065, Japan}
\altaffiltext{12}{Department of Physics, Yamagata University, 
1-4-12 Kojirakawa-machi, Yamagata, Yamagata 990-8560, Japan}
\altaffiltext{13}{National Radio Astronomy Observatory, P.O. Box O,
 Socorro, NM, 87801, USA}
\altaffiltext{14}{Center for Computational Sciences, University of Tsukuba, 1-1-1 Tennodai, Tsukuba
, Ibaraki 305-8577, Japan} 

\email{kitayama@ph.sci.toho-u.ac.jp} 

\maketitle

 \begin{abstract}
  We present measurements of the thermal Sunyaev-Zel'dovich effect (SZE)
  toward SPT-CL J2334-4243 (the Phoenix galaxy cluster) at $z=0.597$ by
  the Atacama Large Millimeter/submillimeter Array (ALMA) in Band 3.
  The SZE is imaged at $5''$ resolution (corresponding to the physical
  scale of 23 $h^{-1}$kpc) within 200 $h^{-1}$kpc from the central
  galaxy with the peak signal-to-noise ratio exceeding 11.  Combined
  with the Chandra X-ray image, the ALMA SZE data further allow for
  non-parametric deprojection of electron temperature, density, and
  entropy. Our method can minimize contamination by the central AGN and
  the X-ray absorbing gas within the cluster, both of which largely
  affect the X-ray spectrum. We find no significant asymmetry or
  disturbance in the SZE image within the current measurement errors.
  The detected SZE signal shows much higher central concentration than
  other distant galaxy clusters and agrees well with the average
  pressure profile of local cool-core clusters. Unlike typical clusters
  at any redshift, gas temperature drops by at least a factor of 5
  toward the center.  We identify $\sim 6 \times 10^{11} M_\odot$ cool
  gas with temperature $\sim 3 $ keV in the inner $20 ~h^{-1}$
  kpc. Taken together, our results imply that the gas is indeed cooling
  efficiently and nearly isobarically down to this radius in the Phoenix
  cluster.
 \end{abstract}

 \section{Introduction}

 It has long been argued that the density of thermal gas particles in
 cores of galaxy clusters is large enough for these particles to cool
 radiatively, leading to a runaway ``cooling flow'' toward the cluster
 center (e.g., \cite{Fabian94} for review).  Such rapid gas cooling or
 associated star formation in central galaxies, however, has not been
 observed in galaxy clusters in the local Universe.  Radiative gas
 cooling must then be suppressed, e.g., by feedback from an Active
 Galactic Nucleus (AGN) often hosted by a central galaxy (e.g.,
 \cite{McNamara07,Fabian12}), while the exact mechanism is yet
 uncertain.
 
 The Phoenix galaxy cluster, SPT-CL J2344-4243, at redshift $z=0.597$
 possibly provides a unique counter example to suppressed gas cooling
 mentioned above. It is the most X-ray luminous galaxy cluster known to
 date with exceptionally high concentration of thermal gas within the
 central 100 kpc; the predicted cooling (or mass deposition) rate
 amounts to $\dot{M}_{\rm cool} = 2000 - 4000 ~M_\odot$ yr$^{-1}$
 \citep{McDonald12,McDonald13,Ueda13}. The central galaxy of the Phoenix
 cluster is also unique in that its star formation rate $\dot{M}_{\rm
 SF} = 400 - 900 ~M_\odot$ yr$^{-1}$ \citep{McDonald12, McDonald13,
 Mittal17} is among the largest in any galaxy at $z<1$, it hosts a dusty
 type II quasar \citep{Ueda13}, and it is associated with extended
 filaments of warm ($\sim 10^4$ K) ionized gas \citep{McDonald14a} and
 cold molecular gas \citep{Russell17}.

 One of the key questions on the nature of the Phoenix cluster is how
 much thermal gas is in fact cooling to sufficiently low temperatures.
 While the presence of cool plasma with temperature $ kT \ltsim 3$ keV,
 where $k$ is the Boltzmann constant, has been reported in the
 literature \citep{Ueda13,Pinto18,McDonald19}, the deposition rate of
 such gas is often inferred to be lower than the values quoted above
 \citep{Tozzi15, Pinto18, McDonald19}. This may imply that radiative
 cooling is still suppressed at $kT > 3$ keV in this cluster.  A major
 challenge in the X-ray analysis of this cluster is the presence of a
 bright AGN and possibly cold neutral gas in the central galaxy; the
 former dominates the X-ray emission at energies $E>2$ keV and the
 latter modifies the spectrum at $E<2$ keV via absorption
 (\cite{McDonald19}; see also section \ref{sec-xray}).  It is hence
 crucial to explore the physical states of the gas by other independent
 methods.
 
 In this paper, we report on measurements of the thermal
 Sunyaev-Zel'dovich effect (SZE: \cite{Sunyaev70,Sunyaev72}) toward the
 Phoenix cluster by the Atacama Large Millimeter/submillimeter Array
 (ALMA). While the cluster was first identified via the SZE by the South
 Pole Telescope (SPT) \citep{Carlstrom11, Williamson11}, its internal
 structures were not explored because of moderate angular resolution
 ($>1'$) of the SPT. We present the first spatially-resolved SZE image
 of the Phoenix cluster at angular resolution of $5''$ using ALMA Band 3
 (section \ref{sec-results}). We further explore thermodynamic
 properties of the hot gas in conjunction with the X-ray data taken by
 Chandra (section \ref{sec-implication}). The high resolution SZE image
 by ALMA provides an independent and complementary probe of the
 intracluster medium (ICM) to X-rays.

Throughout the paper, we adopt a standard set of cosmological density
parameters, $\Omega_{\rm M}=0.3$ and $\Omega_{\rm \Lambda}=0.7$. We use
the dimensionless Hubble constant $h\equiv H_0/(100 \mbox{km/s/Mpc})$;
given controversial results on the value of $h$ (e.g., \cite{Verde19})
we do not fix it unless otherwise stated.  In this cosmology, the
angular size of 1$''$ corresponds to the physical size of 4.67
$h^{-1}$kpc at the source redshift $z=0.597$. The errors are given in
1$\sigma$ and the coordinates are given in J2000.

\begin{table*}
 \caption{Summary of observations.} \label{tab-obs}
  \begin{center}
    \begin{tabular}{ccc}
     \hline
     Array & 12-m & 7-m \\ \hline
     Date & March 17 -- 19, 2016 & May 3 -- June 12, 2016 \\
     Total on-source time [hr] & 3.21 &  8.06  \\
     Number of execution blocks &4 & 15 \\ 
     Number of antennas & $36 - 37$ & $7 - 10$ \\
     Flux calibrator & Neptune & Neptune, Uranus \\
     Phase calibrator & J2336-4115 & J2328-4035 \\
     Bandpass calibrator & J2357-5311& J0006-0623, J2258-2758, J0538-4405 \\
     Central frequency [GHz] & 92 & 92 \\
     Band widths [GHz] & 7.5 & 7.5  \\  
     Baseline coverage [k$\lambda$]
     & $3.7 - 145$ & $2.1 - 15.6$ \\
     Primary beam FWHM at the central frequency [arcsec] & $62$ &  $107$ \\
          Number of pointings & 7  & 7 \\
      \hline
    \end{tabular}
  \end{center}
\end{table*}

\begin{table*}
 \caption{Properties of synthesized images.}\label{tab-image}
  \begin{center}
    \begin{tabular}{ccccc}
      \hline
     Array & 12-m & 12-m ($>30$k$\lambda$ only) &
     7-m & 12-m + 7-m \\ \hline
     Beam major axis FWHM [arcsec] & $2.22$ & $1.86$
	     & $19.7$ & $2.25$ \\ 
     Beam minor axis FWHM [arcsec] & $1.89 $ & $1.61$
	     & $11.5 $ & $1.92$ \\ 
     Beam position angle [deg] & $76.6$ & $74.4$  &$84.2$  &$76.6$  \\ 
Average 1$\sigma$ noise [mJy/beam] & 0.0123 & 0.0144 & 0.0714 & 0.0122 \\ 
      \hline
    \end{tabular}
  \end{center}
\end{table*}

\section{Observations and data reduction}
\label{sec-obs}

SPT-CL~J2344--4243 was observed by the 12-m and 7-m arrays of ALMA
(project code: 2015.1.00894.S) as summarized in table \ref{tab-obs}. The
observations were executed in 4 and 15 separate blocks for the 12-m and
7-m arrays, respectively. Each execution block lasted less than 80
minutes including overheads.  The number of antennas and the calibrators
slightly varied among the execution blocks as listed in table
\ref{tab-obs}. All the data were taken at four continuum bands centered
at 85, 87, 97, and 99 GHz, yielding the overall central frequency of 92
GHz with an effective bandwidth of 7.5 GHz. Compact configurations were
adopted to cover the baseline ranges of 3.7--145 k$\lambda$ and
2.1--15.6 k$\lambda$ for the 12-m and 7-m arrays, respectively, where
$\lambda$ is the observed wavelength. They yielded the angular
resolution and the maximum recoverable scale of $\sim 2''$ and $\sim
60''$, respectively.

The target field, centered at (\timeform{23h44m43.90s},
\timeform{-42D43'12.00''}), had a diameter of
about 1.5$'$ covered with 7 hexagonal mosaic pointings by both
arrays. An equal spacing of 34.2$''$ between the pointings was adopted,
yielding approximately the Nyquist sampling for the 12-m array and much
denser sampling for the 7-m array.


Throughout this paper, we used the visibility data produced by the
second stage of ALMA's Quality Assurance process (QA2).  Imaging was
done with the Common Astronomy Software Applications package (CASA:
\cite{McMullin07}) version 5.6.1.  The procedure was similar to
\citet{Kitayama16}. We adopted the multi-frequency synthesis mode in
joint mosaic imaging.  Natural weighting was used and all the images
presented were corrected for primary beam attenuation.

Table \ref{tab-image} lists the parameters of the synthesized beams as
well as the $1\sigma$ noise levels of the synthesized image within
$45''$ from the field center. The noise levels were measured on a
difference map created after subtracting the compact sources as
described in section \ref{sec-source}, dividing the data set in half,
taking a difference between their dirty images, and dividing it by a
factor 2 to correct for the reduction of the integration time.

\section{Results}
\label{sec-results}

\subsection{Compact sources}
\label{sec-source}

  \begin{table*}[tp]
     \caption{Positions and flux densities of compact sources in the
   observing field. The errors in the positions are less than $0.15''$
   and the effective angular resolution is $1.86''\times 1.61''$ FWHMs
   (table \ref{tab-image}). Sources C1, C3, and W are assumed to be
   point-like.  Source C2 is modeled by an elliptical Gaussian (see text
   for details) but the results of a point source model fit are also
   shown for reference. }
  \begin{center}
    \begin{tabular}{ccccc}
     \hline
     Source ID & model &RA (J2000) & Dec (J2000) & 92 GHz flux density
     [mJy]\\ \hline 
     C1 & point-like & \timeform{23h44m43.905s} &\timeform{-42D43'12.548''}
	     & $1.891 \pm 0.010$ \\
     C2 & Gaussian & \timeform{23h44m43.884s} &\timeform{-42D43'10.633''}
		 & $0.090 \pm 0.017$ \\
     C2 & point-like & \timeform{23h44m43.887s} &\timeform{-42D43'10.684''}
	     & $0.068 \pm 0.010$ \\     
     C3 & point-like & \timeform{23h44m43.973s} &\timeform{-42D43'13.493''}
		 & $0.091 \pm 0.010$ \\
     W & point-like & \timeform{23h44m41.661s} &\timeform{-42D43'22.139''}
	     & $0.158 \pm 0.010$ \\     
     \hline
    \end{tabular}
  \end{center}
 \label{tab-source}
  \end{table*}

  \begin{figure*}
   \begin{minipage}{5cm}
    \begin{center}
     \includegraphics[width=4.5cm]{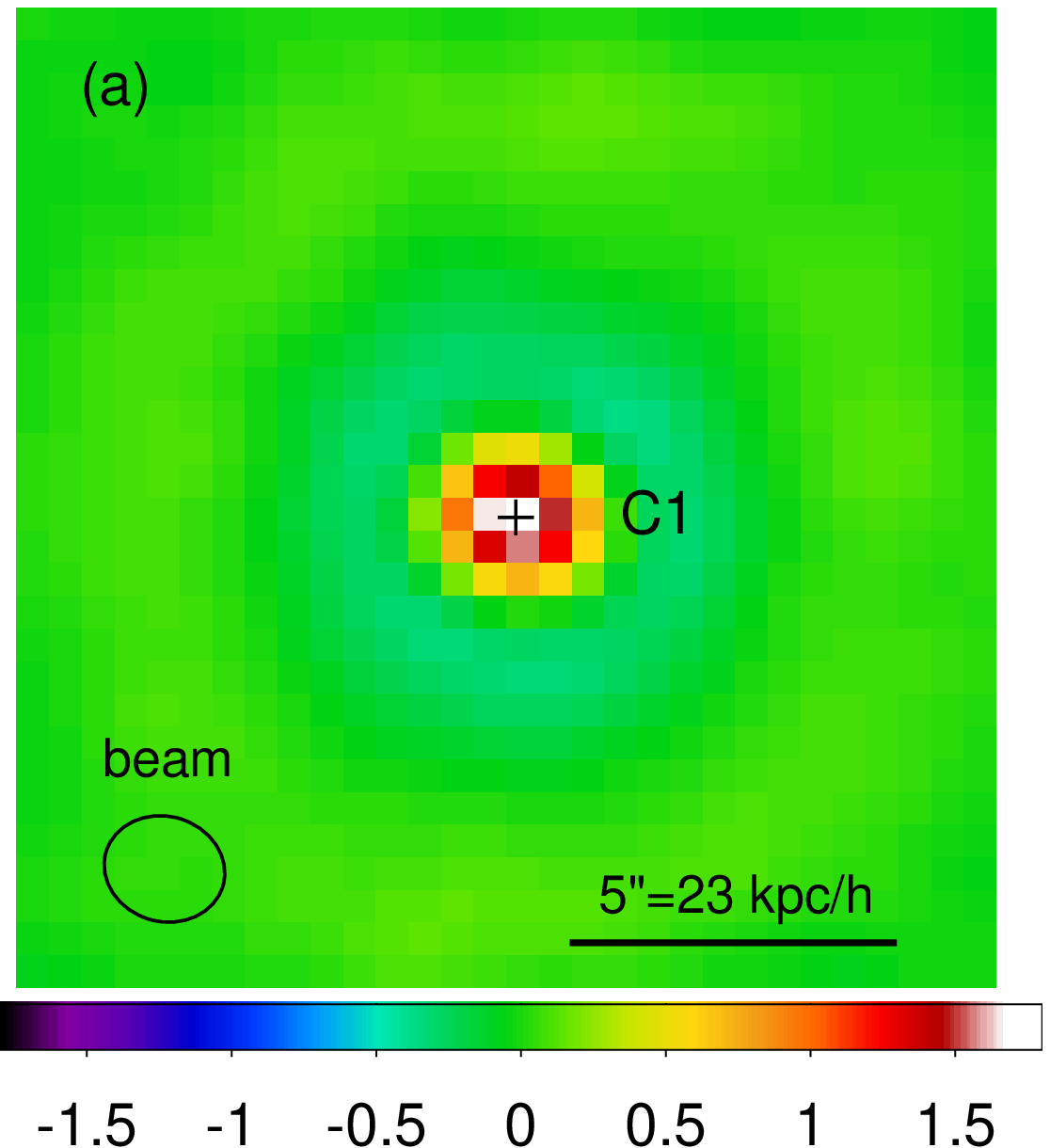}
          \includegraphics[width=4.5cm]{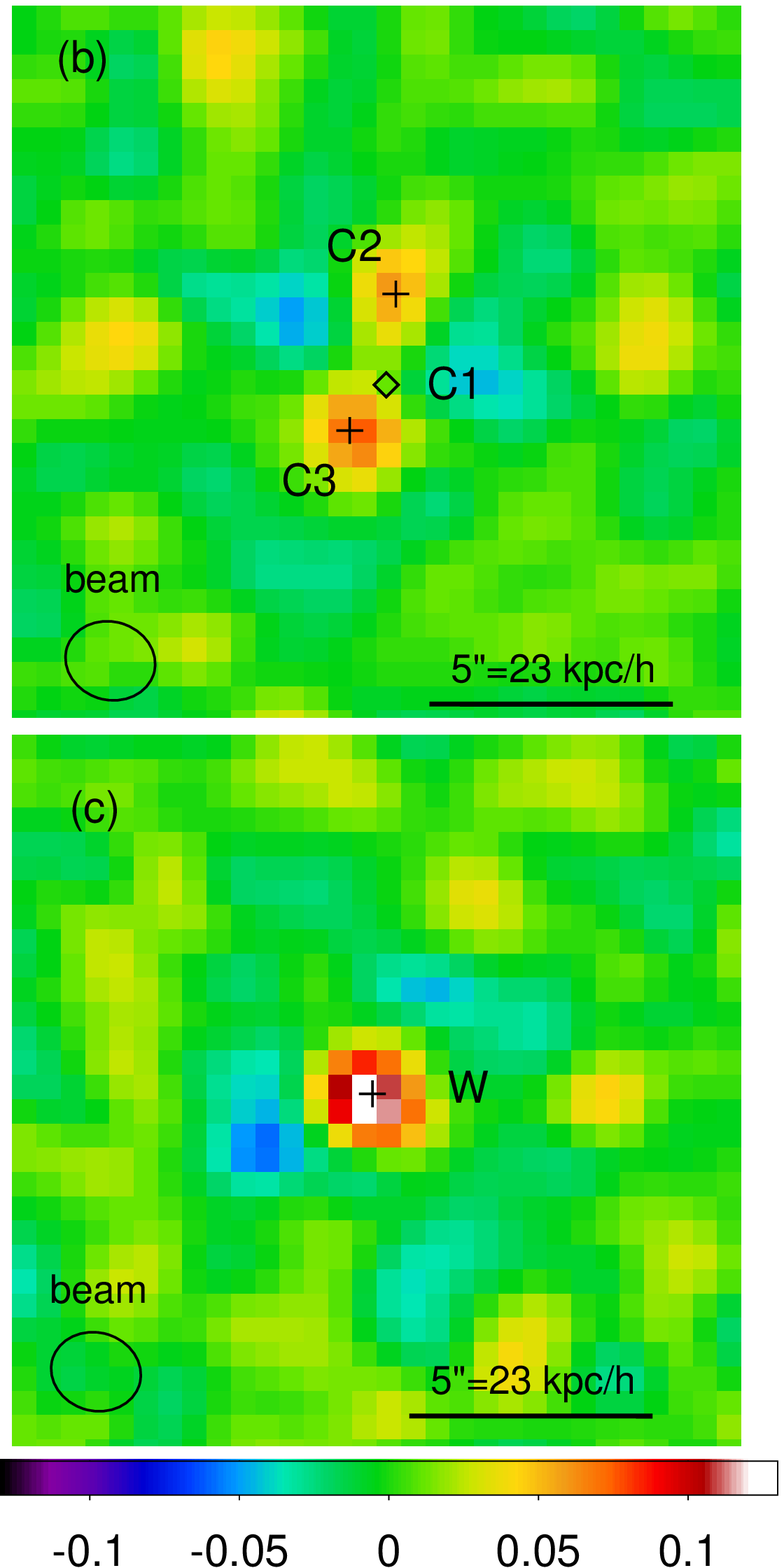}     
    \end{center}
   \end{minipage}
  \begin{minipage}{11.5cm}
	\begin{center}
	 \includegraphics[width=11.5cm]{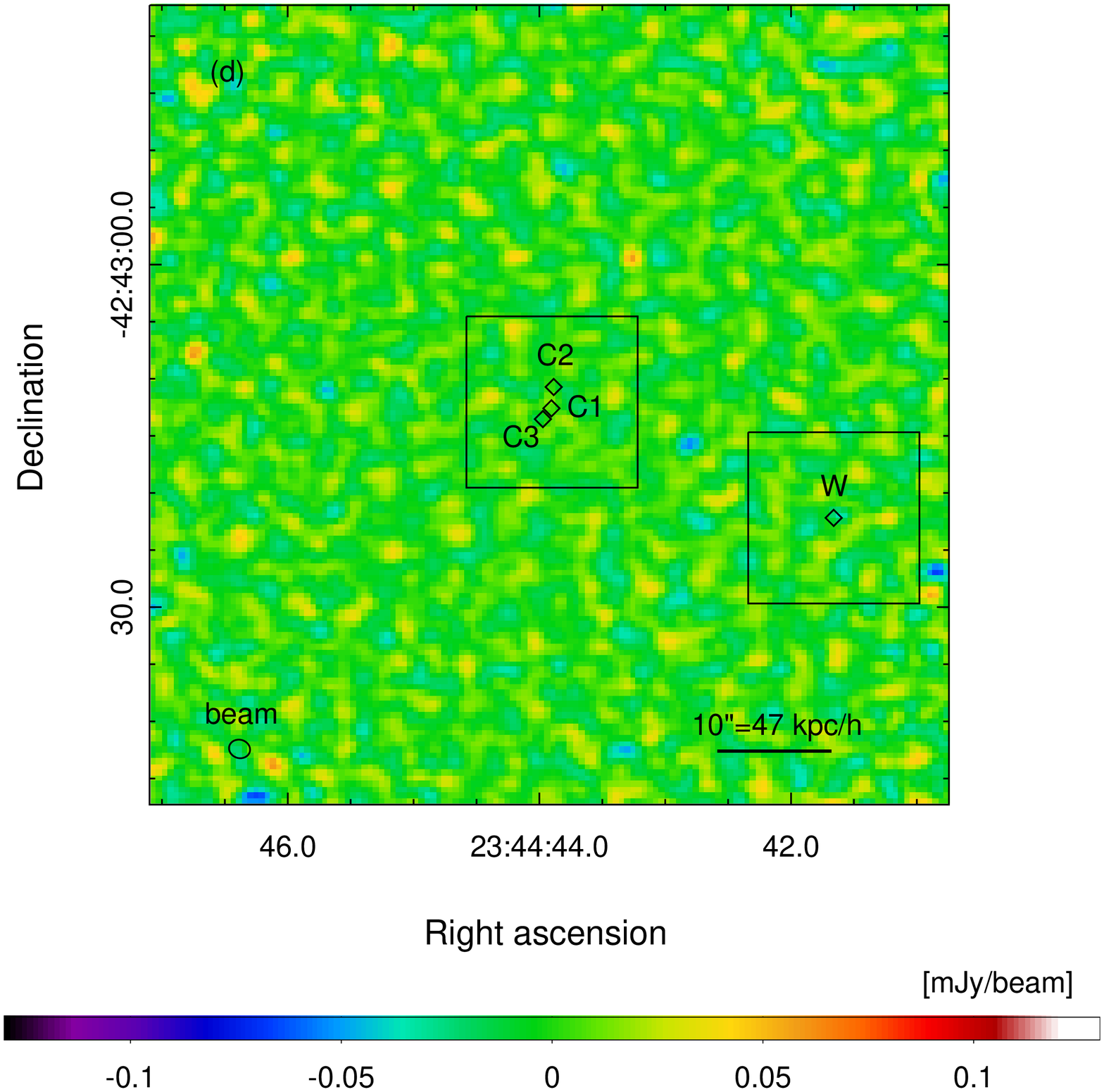}
    \end{center}	 
  \end{minipage}   
   \vspace*{4mm}
   \caption{Dirty images produced by long-baseline ($>30$k$\lambda$)
   visibility data.  Positions of identified sources before and after
   subtraction are marked by crosses and diamonds, respectively.  The
   synthesized beam shape (table \ref{tab-image}) is shown at the
   bottom-left of each panel.  (a) The central $15''\times 15''$ region
   before the sources are subtracted. Sources C2 and C3 are not
   prominent because source C1 is much brighter.  (b) Same as panel a
   but after source C1 is subtracted. (c) The region around source W.
   (d) The image after sources C1, C2, C3, and W are subtracted. Boxes
   indicate the regions shown in panels a, b, and c.}
   \label{fig-source}
  \end{figure*}

There are four compact sources detected by emission above the $5\sigma$
significance level in our target field.  As described below, we
determined their positions and flux densities by fitting with the CASA
task {\it uvmodelfit} the 12-m array visibility data at baselines longer
than 30k$\lambda$ (corresponding to the spatial scale smaller than
$\lambda/(\mbox{30k}\lambda) \sim 7''$) to eliminate contamination by
the extended SZE. Source identification and subtraction are done in the
$uv$ plane and fully independent of the image synthesis method. The
results are summarized in table \ref{tab-source}. These sources are also
detected at 18 GHz by ATCA and their properties are discussed in detail
in a separate paper \citep{Akahori19}.  The assigned source names are
the same as in \citet{Akahori19}; C1, C2, and C3 are central sources in
descending order of their 18GHz flux, whereas W is a western off-center
source.

The brightest central source, C1, is an AGN hosted by the central
galaxy. Figure \ref{fig-source}a shows a dirty image toward source C1
produced by the visibility data at $>30$k$\lambda$.  Throughout this
paper, we denote the projected angular distance and the deprojected
physical distance from source C1 by $\theta$ and $r$, respectively.

When the best-fit point source model for source C1 (table
\ref{tab-source}) is subtracted from the visibility data, two weaker
sources, C2 and C3, become apparent near the cluster center (figure
\ref{fig-source}b). In addition, there is another off-center source, W,
at $\theta \sim 25''$ to the west from source C1 (figure
\ref{fig-source}c). As the shapes of sources C3 and W are consistent
with the synthesized beam, they are modeled by point sources. We checked
that the results in table \ref{tab-source} are essentially unchanged
when these sources are modeled by a Gaussian varying its full width at
half maximum (FWHM) as a free parameter. On the other hand, source C2
appears to be elongated from north-west to south-east. We thus modeled
this source by an elliptical Gaussian, fixing the major-to-minor axis
ratio at 0.05, to obtain the major axis FWHM of $1.3''\pm 0.5''$ and the
position angle of $-9.4^\circ \pm 21.8^\circ$, with the centroid
position and flux density listed in table \ref{tab-source}. We checked
that the results are insensitive to the assumed value of the axis ratio
as long as the minor axis remains negligible compared to the synthesized
beam size. For comparison, if source C2 is modeled by a point source,
the fitted flux is lower by about 25\% (table \ref{tab-source}).

Figure \ref{fig-source}d shows a long baseline ($>30$k$\lambda$) image
after all the detected sources are removed from the visibility data.
The residuals have the root mean square (rms) value of 0.0146 mJy/beam
and are consistent with noise. This confirms that identification and
subtraction of the sources are properly done.

\subsection{The Sunyaev-Zel'dovich effect}
\label{sec-sz}

\begin{figure}[t]
 \begin{center}
  \includegraphics[width=8.3cm]{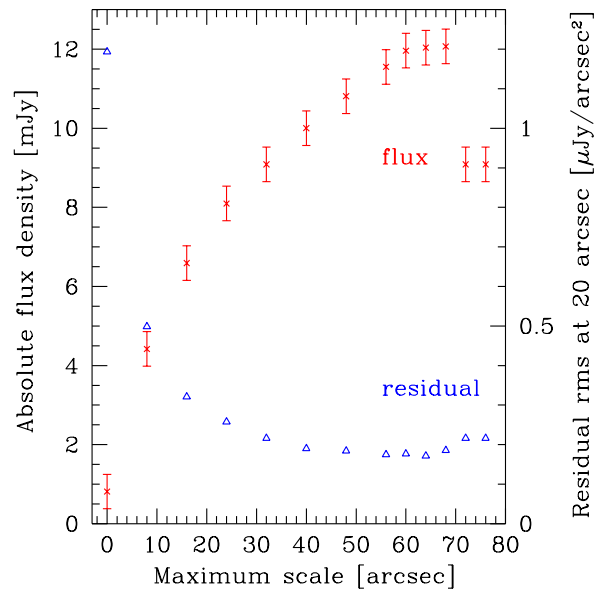} 
 \end{center}
 \caption{Recovered flux and residuals within the field of view ($1.5'$
 in diameter) versus the maximum scale of Gaussians used in Multi-Scale
 CLEAN.  Wherever smaller than the scale shown in the figure, Gaussians
 with FWHMs of $0$, $4''$, $8''$, $16''$, and $32''$ are also used in the
 deconvolution; the smaller scale values are fixed for comparison. The
 plotted residual is the rms value measured on a residual map after
 being corrected for primary beam attenuation and smoothed by a Gaussian
 kernel with $20''$ FWHM.  } \label{fig-scale}
\end{figure}

The source-subtracted visibility data were deconvolved with the
Multi-Scale CLEAN algorithm \citep{Cornwell08,Rich08,Steeb19} using the
CASA task {\it tclean}. We adopted [0, $4''$, $8''$, $16''$, $32''$,
$64''$] as the FWHMs of the Gaussian components used in Multi-Scale
CLEAN implemented in CASA version 5.6.1 \footnote{The algorithm of
Multi-Scale CLEAN has been modified since CASA version 5.6.0
\citep{Steeb19}. We checked that the results of the present paper remain
essentially unchanged by this modification as long as the Gaussian
components are chosen to give maximal recovered flux and minimal
residuals as described in the text.}.  As illustrated in figure
\ref{fig-scale}, this is an optimal choice for the present target and
assure maximal recovered flux as well as minimal residuals. We used a
circular mask region with radius $\theta = 42''$, a flux threshold of
0.024 mJy, and a loop gain of 0.05.

\begin{figure}[t]
   \begin{center}
  \vspace*{-4mm}    
  \includegraphics[width=8.3cm]{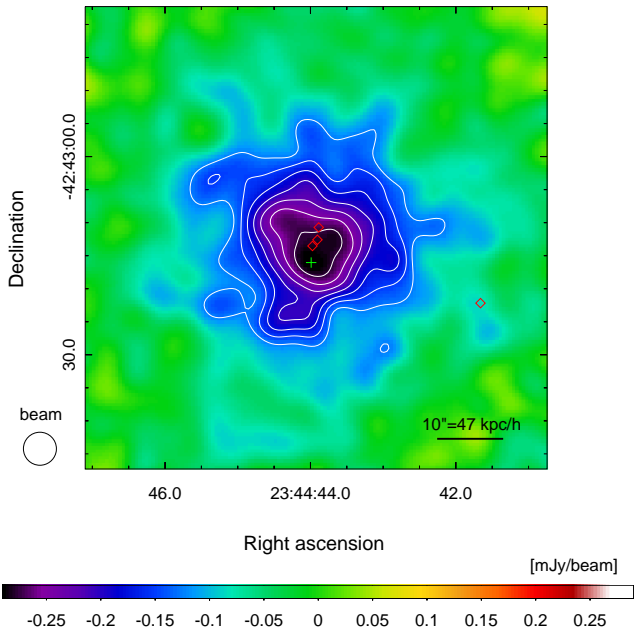} 
   \end{center}
 \caption{ALMA SZE image of the Phoenix cluster at the central frequency
 of 92 GHz smoothed to have a beam size of $5''$ FWHM. Contours show the
 statistical significance of $5-11 \sigma$ in increments of $1 \sigma=
 0.025$ mJy/beam. The positions of the SZE peak and subtracted sources
 are denoted by a cross and diamonds, respectively.}  \label{fig-szmap}
 \end{figure}

Figure \ref{fig-szmap} shows the deconvolved SZE image created from the
visibility data taken by the 12-m and 7-m arrays after the compact
sources described in section \ref{sec-source} are removed. The image has
been smoothed to an effective beam size of $5''$ FWHM for display
purposes; unless otherwise stated, quantitative analysis in this paper
was done on the unsmoothed image with $2.25'' \times 1.92''$ FWHMs.  The
rms noise level measured on the difference map smoothed to the $5''$
resolution is 0.025 mJy/beam at $\theta < 45''$. The SZE decrement is
detected at more than $11\sigma$ statistical significance. The SZE peak
is located at $3.6''$ south from the central AGN (source C1), with the
SZE intensities at the two positions of $-0.294 \pm 0.025$ mJy/beam (SZE
peak) and $-0.280 \pm 0.025$ mJy/beam (source C1 location),
respectively, on the smoothed image. The offset is hence not
statistically significant given the noise level of the ALMA data.  The
integrated SZE flux density within $\theta = 40''$ is $-11.9\pm 0.4$
mJy. The mean signal within the annulus at $40''< \theta < 45''$ is
consistent with zero and $-0.13 \pm 0.17 ~\mu$Jy/arcsec$^2$.

\begin{figure}
 \begin{center}
  \includegraphics[width=8.3cm]{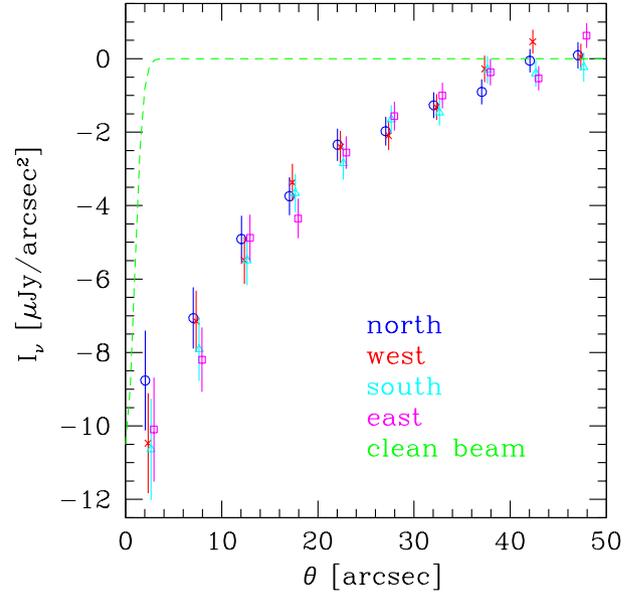} 
 \end{center}
 \caption{Azimuthally averaged SZE intensity profiles as a function of
 the projected distance from the central AGN in the four quadrants;
 north (circles), west (crosses), south (triangles), and east
 (squares). For clarity, symbols are slightly shifted
 horizontally. Dashed line shows an azimuthally averaged shape of the
 synthesized beam.}  \label{fig-radprof}
\end{figure}

We plot in figure \ref{fig-radprof} azimuthally averaged intensity
profiles in four quadrants with position angles of $315^{\circ} -
45^{\circ}$ (north), $45^{\circ} - 135^{\circ}$ (east), $135^{\circ} -
225^{\circ}$ (south), and $225^{\circ} - 315^{\circ}$ (west). The
statistical error in each bin is computed using equation (1) of
\citet{Kitayama16}. The SZE intensities in four quadrants are consistent
with one another within the error, while the signal in the north
quadrant tends to be weaker than the other directions at $\theta <
15''$.

\section{Interpretation and implications}
\label{sec-implication}

\subsection{Comparison to X-ray data}
\label{sec-xray}

To compare with the ALMA SZE data, we extracted the X-ray data of the
Phoenix cluster taken by Chandra ACIS-I (ObsID: 13401, 16135, 16545,
19581, 19582, 19583, 20630, 20631, 20634, 20635, 20636, and 2079). The
total exposure time is 551.5 ksec. The data were processed using CIAO
version 4.11 \citep{Fruscione06} and the Calibration database (CALDB)
version 4.8.2.  The backgrounds were estimated from the off-center
region at $3' < r < 5'$ from the central AGN, where the ICM emission is
negligible. We used the data at observed energies $E=0.7-7.0$ keV to
minimize the effects of the ACIS contamination layer and the
instrumental background (e.g., \cite{Bartalucci14, Plucinsky18}).
Throughout the analysis, we assumed that the ICM is in collisional
ionization equilibrium, the abundance ratio of elements heavier than
helium is that of \citet{Anders89}, and the Galactic hydrogen column
density toward the Phoenix cluster is $1.52 \times 10^{20}$ cm$^{-2}$
\citep{Kalberla05}. We fixed the helium mass fraction at $Y=0.25$, which
is nearly unchanged between the primordial gas and the solar photosphere
(e.g., \cite{Asplund09,Planck18}).  Spectral fitting was done with XSPEC
version 12.10.0e \citep{Arnaud96}.

\begin{figure*}[t]
 \begin{center}
  \includegraphics[width=8.4cm]{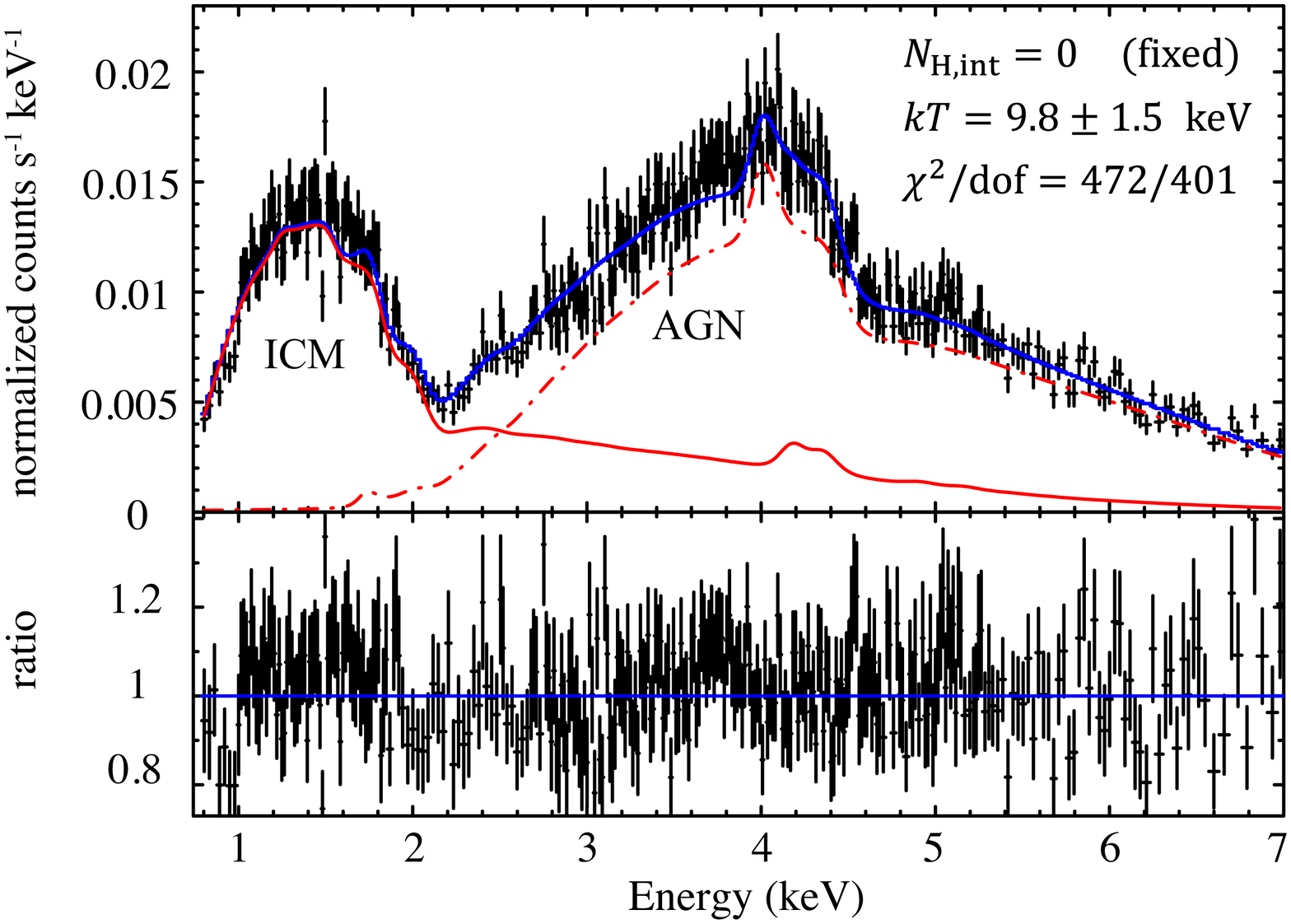}
  \includegraphics[width=8.4cm]{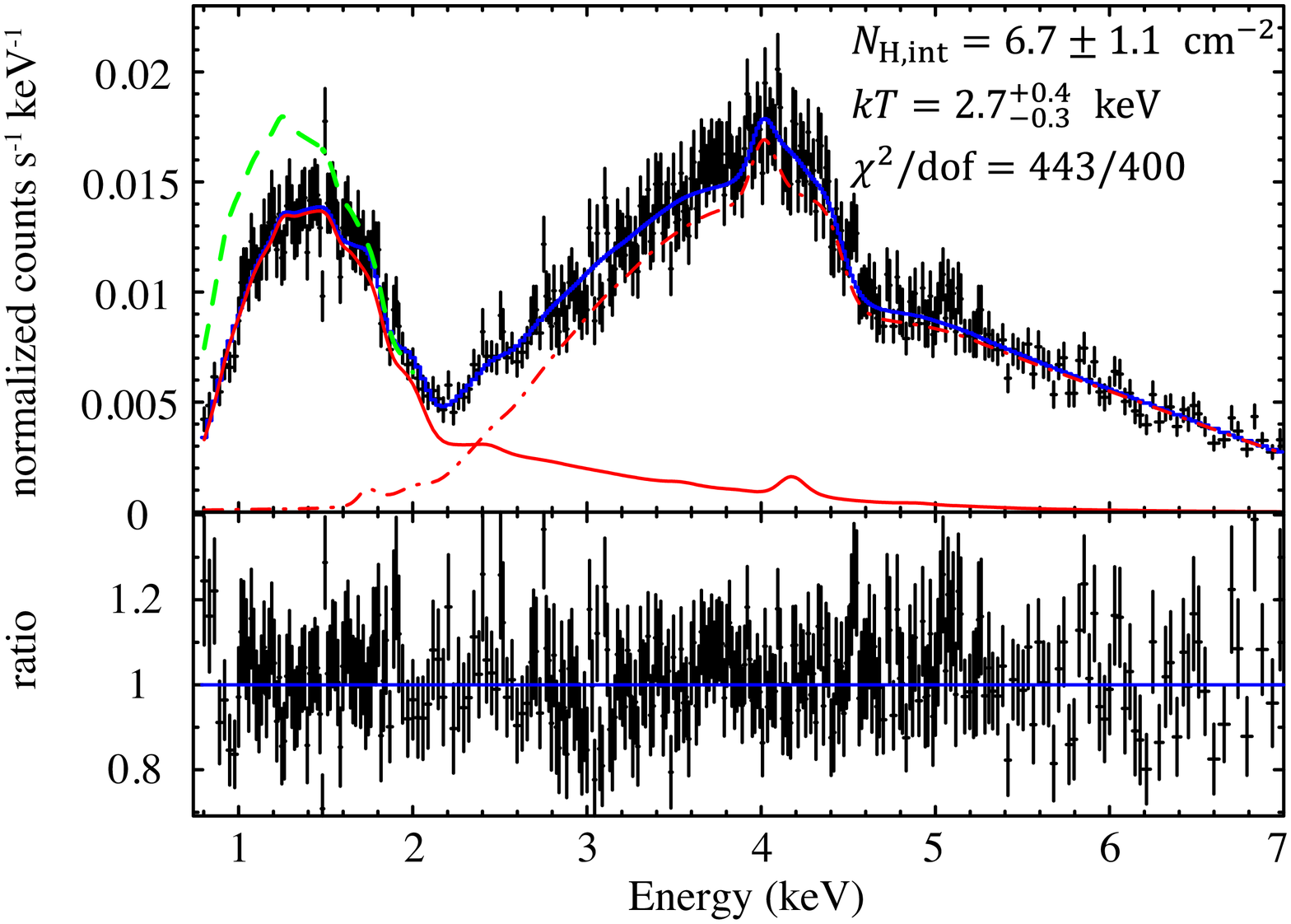}   
 \end{center}
 \caption{Chandra X-ray spectrum of the Phoenix cluster at $\theta_{\rm
 X}<1.5''$ (top) and the residuals with respect to the best-fit models
 (bottom). {\it Left:} Lines show the best-fit models of the ICM without
 intrinsic absorption (red solid), the central AGN (red dot-dashed), and
 their sum (blue solid). {\it Right:} Same as the left panel except that
 intrinsic absorption within the cluster is included in the fit.  For
 reference, a green dashed line shows the ICM spectrum at $E<2$ keV free
 from intrinsic absorption.  }  \label{fig-xspec}
\end{figure*}

There is an X-ray bright type II quasar near the center of the Phoenix
cluster \citep{Ueda13, McDonald15, McDonald19}.  The $2.0-7.0$ keV
brightness has a prominent point-like peak at (\timeform{23h44m43.962s},
\timeform{-42D43'12.412''}) and we refer to this position as ``the X-ray
center'' of this cluster; it agrees with the position of the central AGN
(source C1) in the ALMA 92 GHz map within $0.6''$. For definiteness, we
denote the projected distance from the X-ray center by $\theta_{\rm
X}$. The emission around the X-ray center is dominated by the quasar and
the ICM at $E>2$ keV and $E<2$ keV, respectively (figure
\ref{fig-xspec}). We modeled the spectrum of the former by an obscured
(by a torus) power-law plus an iron fluorescent line at the rest-frame
energy of 6.4 keV as in \citet{Ueda13}.  Following \citet{McDonald19},
we also took into account additional photoelectric absorption by cool
gas within the Phoenix cluster (see the next paragraph for details).
Unless otherwise stated, there were in total five position-dependent
free parameters in our spectral model; projected temperature of the ICM,
projected metallicity of the ICM, column density of an intrinsic
absorber within the cluster, spectral normalization factors for the ICM
and for the central AGN. In addition, three parameters of the central
AGN (the spectral index, column density of an obscuring torus, and the
flux ratio between the 6.4 keV line and the power-law continuum) were
varied when fitting the spectrum at $\theta_{\rm X} < 1.5''$ and fixed at
their best-fit values elsewhere.

Figure \ref{fig-xspec} explicitly shows the impact of intrinsic
absorption within the Phoenix cluster mentioned above.  If intrinsic
absorption is not taken into account, the projected ICM temperature at
$\theta_{\rm X} < 1.5''$ is $kT=9.8 \pm 1.5$ keV with $\chi^2/{\rm dof}
= 472/401$ (left panel), where dof denotes a degree of freedom of the
fit.  Inclusion of intrinsic absorption significantly decreases the
temperature to $kT=2.7^{+0.4}_{-0.3}$ keV and improves the fit to
$\chi^2/{\rm dof} = 443/400$ (right panel); the best-fit hydrogen column
density of the absorber $N_{\rm H, int}= (6.7 \pm 1.1) \times 10^{21}$
cm$^{-2}$ is also consistent with figure 6 of \citet{McDonald19}.  In
other words, intrinsic absorption largely modifies the spectral shape at
0.7--2.0 keV and leads to a reduction of the best-fit temperature by
more than a factor of 3. This reflects the fact that the ICM continuum
is overwhelmed by the AGN at $E>2$ keV and the ICM metal lines are
sensitive to intrinsic absorption at $E<2$ keV.  It is hence meaningful
to perform an independent measurement of the ICM temperature without
relying on the X-ray spectrum. We will discuss such a complementary
probe using the SZE data in section \ref{sec-deproj}.

As in \citet{Kitayama16}, we performed X-ray thermodynamic mapping using
the contour binning algorithm \citep{Sanders06}. In addition to the
central region at $\theta_{\rm X} < 1.5''$ mentioned above, we defined
subregions with nearly equal photon counts by adopting the
signal-to-noise ratio (S/N) threshold of 100 (i.e., $\sim 10000$ counts)
in the $0.7-7.0$ keV band. The $0.7-7.0$ keV spectrum in each subregion
was fitted by the model mentioned above. Typical statistical errors are
25 \% and 5 \% for the temperature and the density of the ICM,
respectively; the errors are mainly driven by the limited energy range,
particularly near the central AGN, available for determining the ICM
properties.  We also checked that the AGN emission is consistent with
the point spread function (PSF) of ACIS-I and becomes negligible at
$\theta_{\rm X} > 5''$; we excluded the AGN emission in the spectral
fitting at such distances.  The absorption-corrected intrinsic AGN
luminosity is $L_{\rm X}(0.7-7.0 \mbox{ keV}) = (1.81 \pm 0.09) \times
10^{45} h^{-2}$ erg s$^{-1}$, which amounts to $\sim 40 \%$ of the X-ray
luminosity from the entire ICM.

\begin{figure*}[tp]
 \begin{center}
  \includegraphics[width=8.4cm]{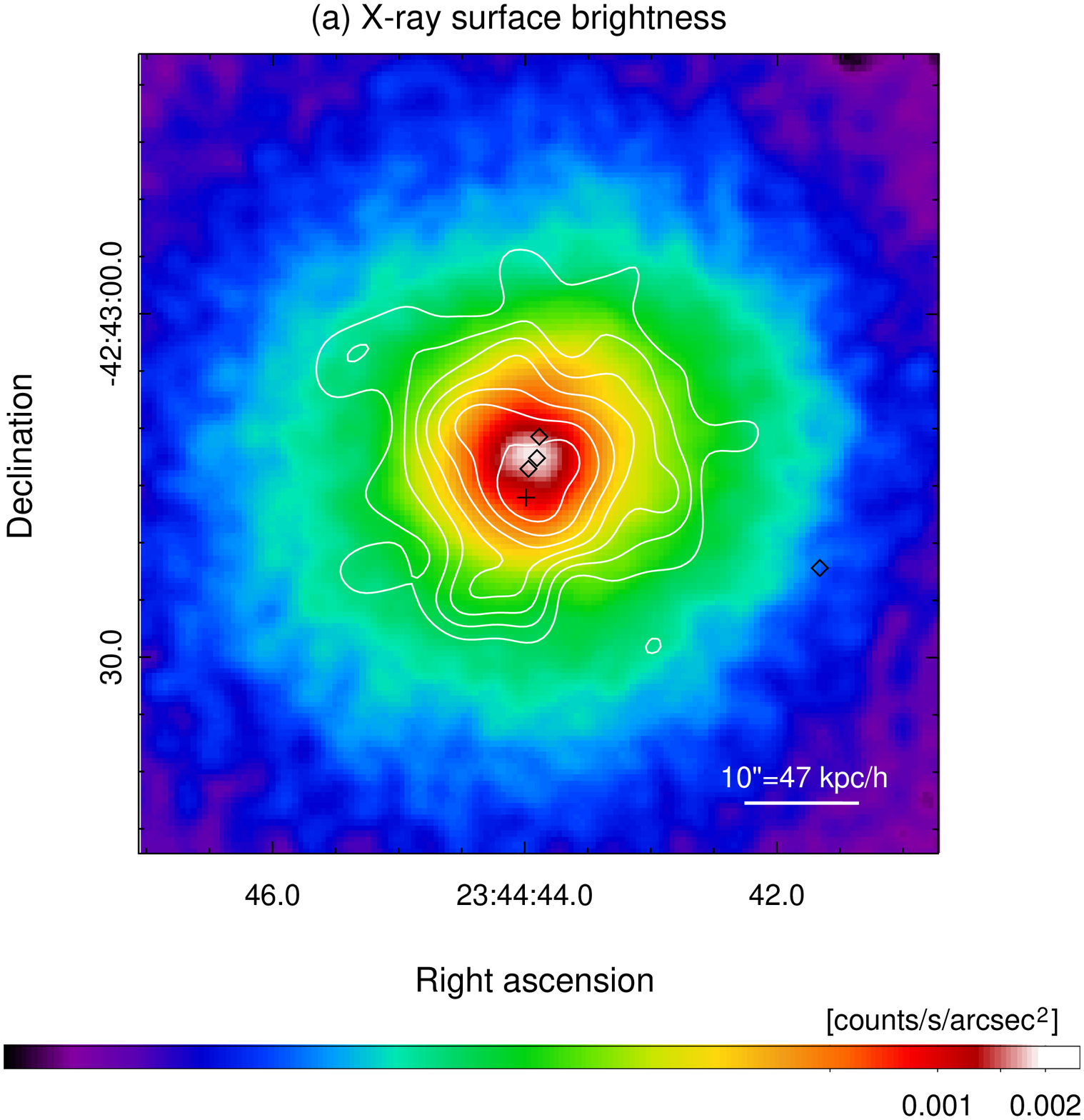} 
  \includegraphics[width=8.4cm]{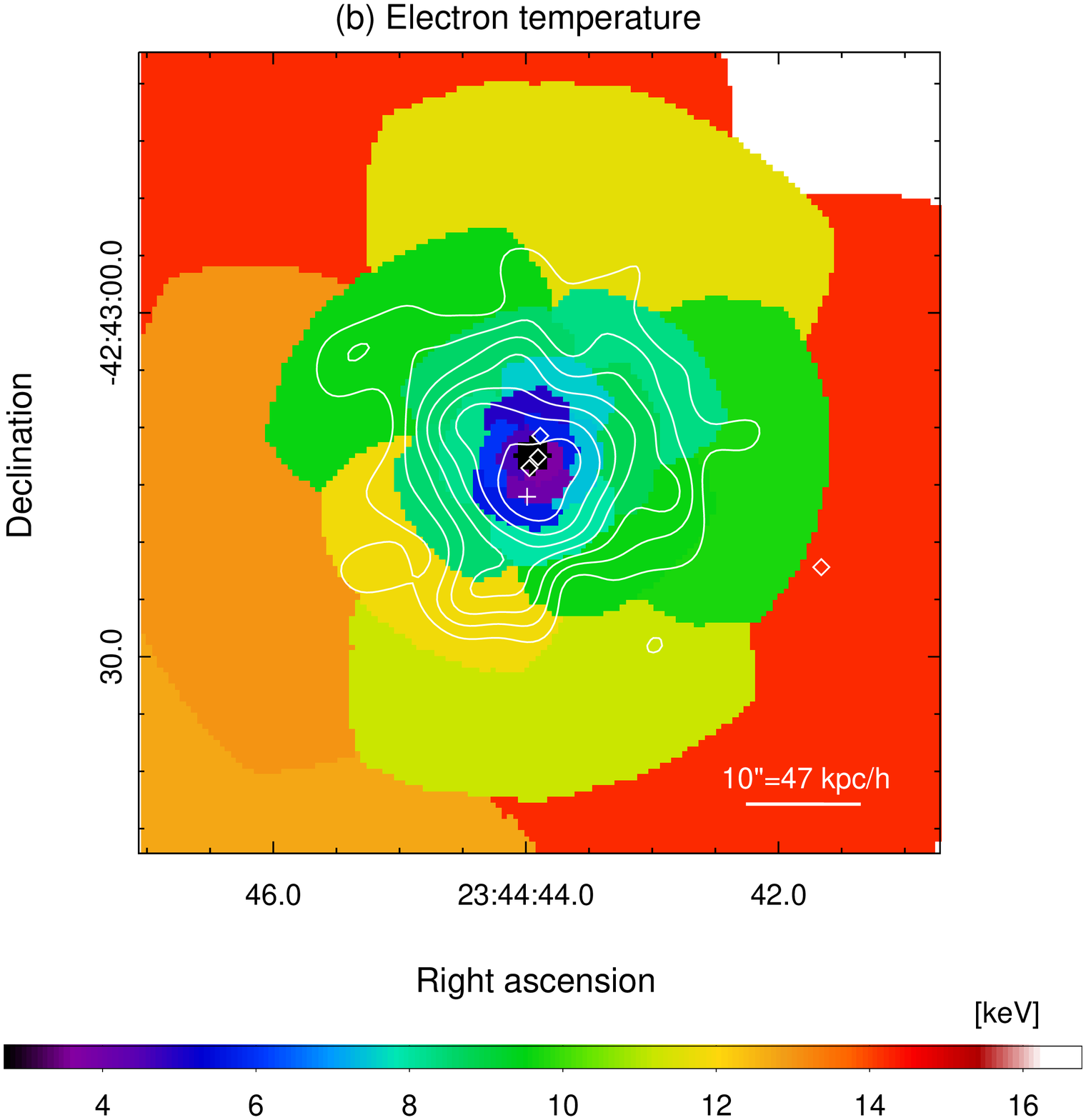} 
  \includegraphics[width=8.4cm]{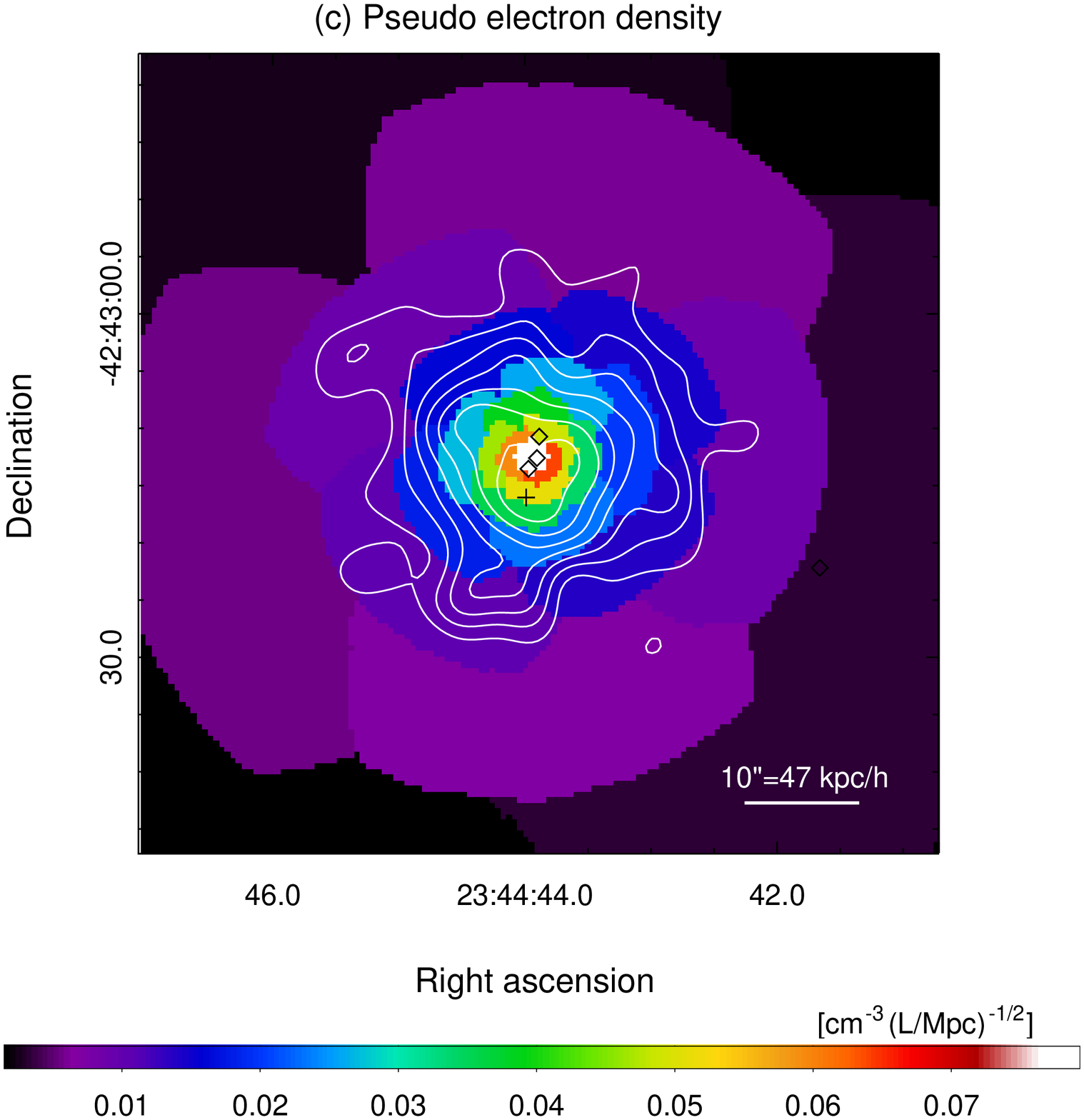} 
  \includegraphics[width=8.4cm]{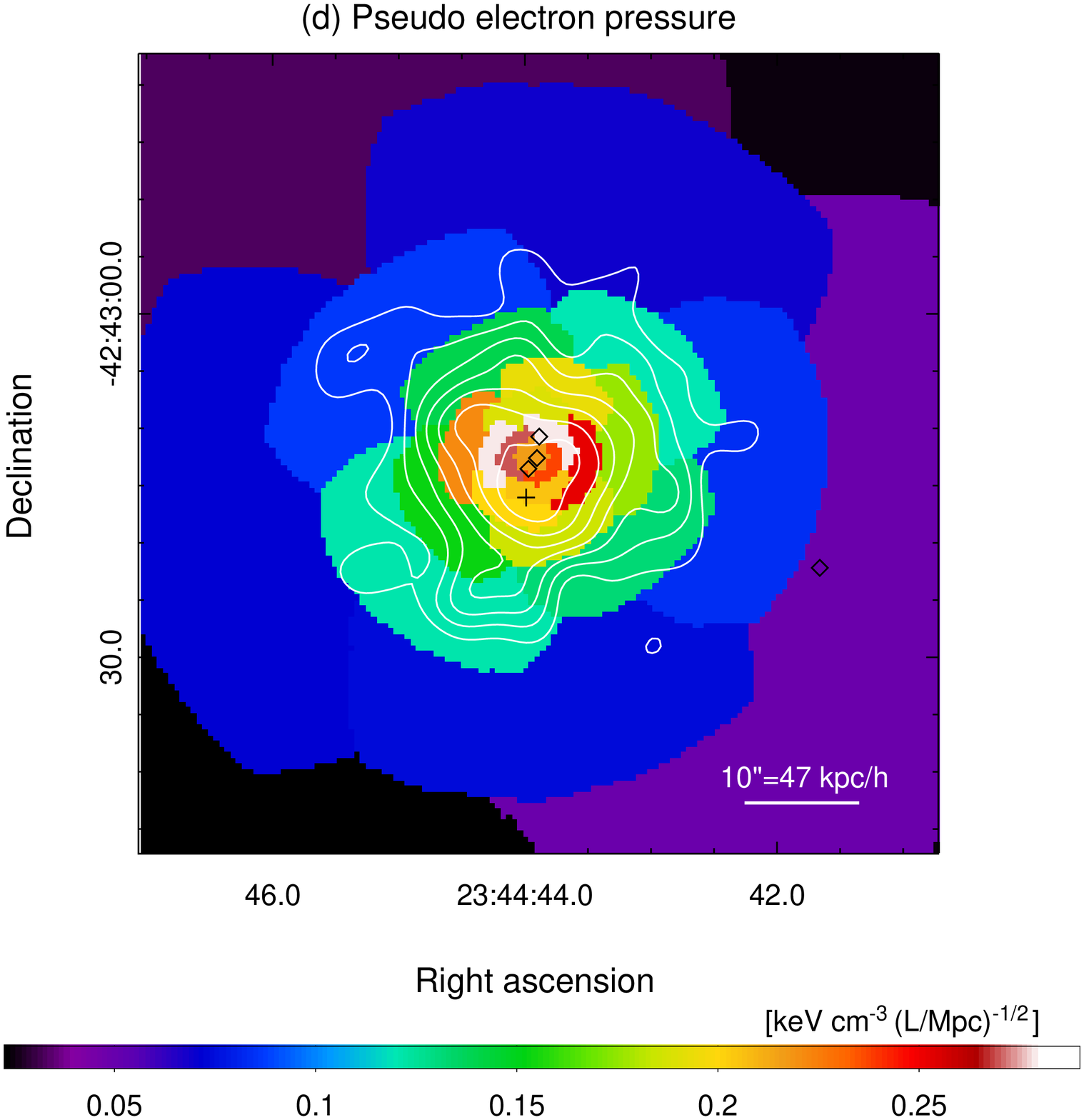} 
 \end{center}
 \caption{Chandra X-ray maps of the Phoenix cluster. Contours show the
 significance levels of the ALMA SZE image plotted in figure
 \ref{fig-szmap}, and a cross and diamonds indicate the positions of the
 SZE peak and subtracted sources, respectively.  Subregions in panels b,
 c, and d are defined by the contour binning algorithm
 \citep{Sanders06}.  (a) X-ray surface brightness in the 0.7--2.0 keV
 band in counts s$^{-1}$ arcsec$^{-2}$, smoothed by a Gaussian kernel
 with $2.3''$ FWHM. The color is shown in a logarithmic scale. (b)
 Projected X-ray spectroscopic temperature in keV. (c) Pseudo electron
 density in cm$^{-3} \times (L/{\rm Mpc})^{-1/2}$ assuming a uniform
 line-of-sight depth of $L$. (d) Pseudo electron pressure in keV
 cm$^{-3} \times (L/{\rm Mpc})^{-1/2}$. } \label{fig-xray}
\end{figure*}

\begin{figure*}[tp]
 \begin{center}
  \includegraphics[width=8.4cm]{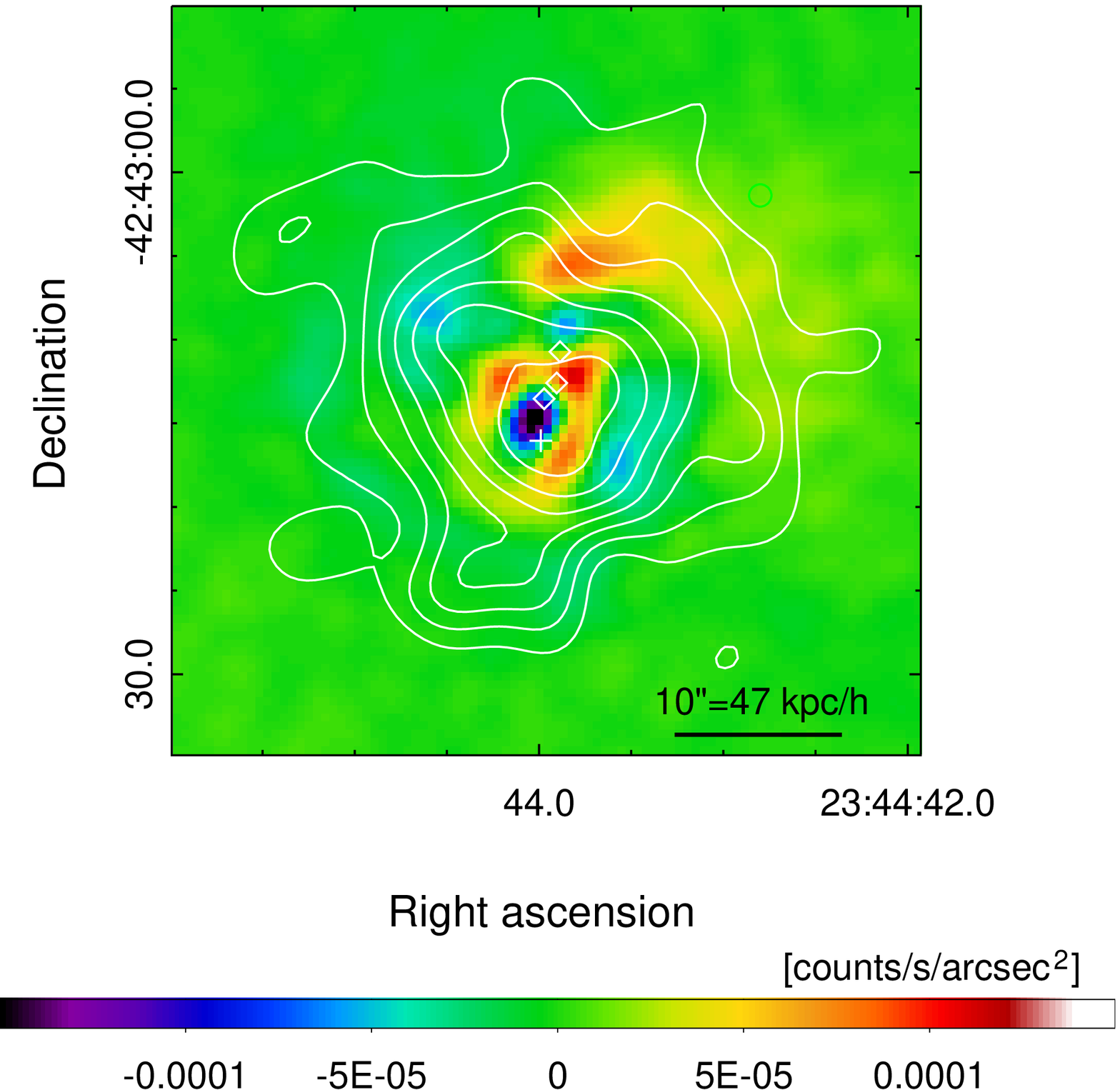}
    \includegraphics[width=8.4cm]{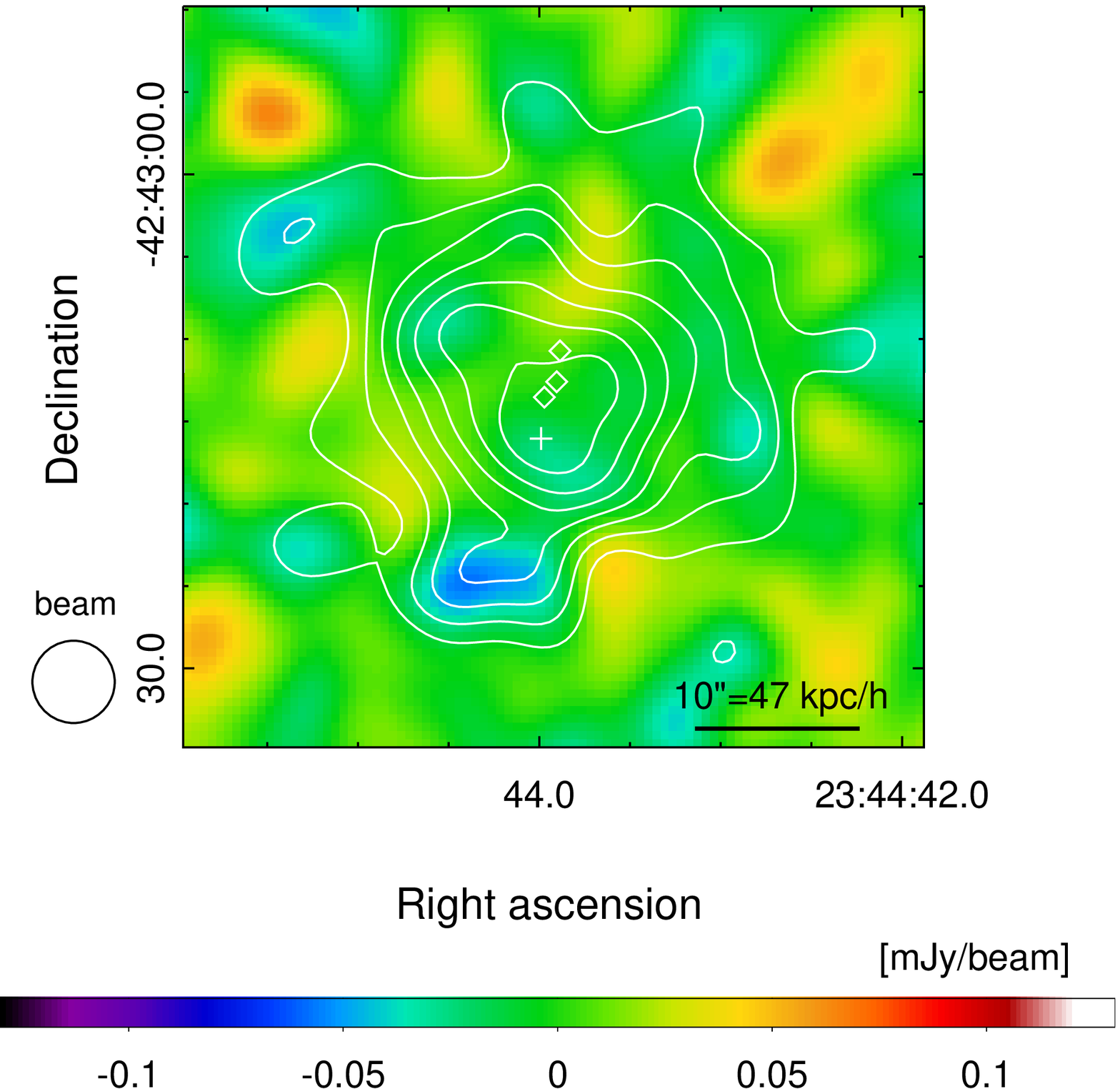} 
 \end{center}
 \caption{Residual X-ray and SZE images of the Phoenix cluster after
 subtracting the mean signal as described in the text.  {\it Left}:
 Chandra X-ray surface brightness in the 0.7--2 keV band, smoothed by a
 Gaussian kernel with $2.3''$ FWHM.  {\it Right}: ALMA SZE brightness at
 92 GHz smoothed to $5''$ FWHM with the rms noise of $0.025$
 mJy/beam. In both panels, contours show the significance levels of the
 ALMA SZE image plotted in figure \ref{fig-szmap}, and a cross and
 diamonds indicate the positions of the SZE peak and subtracted sources,
 respectively.}  \label{fig-res}
\end{figure*}

Figure \ref{fig-xray} shows X-ray measured quantities overlaid with the
ALMA SZE contours. The $0.7-2.0$ keV surface brightness (panel a) traces
the line-of-sight integral of density squared of the ICM, which is
highly concentrated around the central AGN.  On the other hand,
projected temperature (panel b) drops to $\sim 3$ keV near the center
and is lower in the north-south direction.  Pseudo electron density
assuming a uniform line-of-sight depth of $L$ (panel c) tends to be
higher in the north-south direction across the center. Consequently,
pseudo electron pressure (panel d), a product of the quantities plotted
in panels b and c, is less concentrated than the density; it decreases
only by a factor of $\sim 2$ from the center to $\theta \sim 15''$. Note
that absolute values of the pseudo pressure are arbitrary ($\propto
L^{-1/2}$) and the plotted values are also subject to statistical errors
of $\sim 25\%$.  Overall trend of the pseudo pressure map still shows
qualitative agreement with the ALMA SZE image.

To further examine the departure from symmetry in the observed X-ray and
SZE data, we plot in figure \ref{fig-res} residual images after
subtracting the mean signal in a model-independent manner. For this
purpose, we first searched for an ellipse that minimizes the variance of
the $0.7-2.0$ keV X-ray brightness relative to its mean at
$\bar{\theta}_{\rm X}<5''$, where $\bar{\theta}_{\rm X}$ is the
geometrical mean of semi-major and semi-minor axis lengths around the
X-ray center. We found that such an ellipse has the axis ratio of 0.97,
the position angle of $-112^{\circ}$, and the center at ($\Delta$R.A.,
$\Delta$Dec)= ($0.1''$, $-0.2''$) from the X-ray center.  We then
computed the mean X-ray brightness over this ellipse as a function of
$\bar{\theta}_{\rm X}$ and subtracted it from the X-ray brightness at
each sky position. We also subtracted from the SZE image the mean SZE
brightness over the same ellipse as used for the X-ray image.

Figure \ref{fig-res} confirms the presence of X-ray cavities and
overdense regions around the central AGN reported previously
\citep{McDonald15, McDonald19}. It further shows the absence of
significant disturbance in the residual SZE image above the noise level
($1 \sigma = 0.025$ mJy/beam at $5''$ FWHM). This suggests that the
ICM in this region is consistent with being isobaric.

To be more quantitative, we applied the same method as in section 3.4 of
\citet{Ueda18} and inferred the equation of state of observed
perturbations. Within $10''$ from the X-ray center and assuming the
line-of-sight depth of $l=100$ kpc, the X-ray data give $|\Delta I_{\rm
X}|/\langle I_{\rm X} \rangle = 0.115$, $kT = 5.9 \pm 0.3$ keV, and
$\sqrt{\langle n_{\rm e}^2\rangle} = (0.102 \pm 0.001)(l/100 \mbox{
kpc})^{-1/2}$ cm$^{-3}$, where $I_{\rm X}$ is the X-ray surface
brightness and $n_{\rm e}$ is the electron number density.  Using
equation (4) of \citet{Ueda18}, we obtained the mass density
perturbation of $\Delta \rho = (1.14 \pm 0.02) \times 10^{-26} (l/100
\mbox{ kpc})^{-1/2}$ g cm$^{-3}$. For the same region, the SZE data give
1$\sigma$ upper limit on the pressure perturbation of $\Delta p < 1.17
\times 10^{-10} (l/100 \mbox{ kpc})^{-1}$ erg cm$^{-3}$. They lead to an
upper limit on $w \equiv \Delta p/\Delta \rho$ of $\sqrt{w} < 1010
(l/100 \mbox{ kpc})^{-1/4}$ km s$^{-1}$. While the limit is still weak,
the observed perturbations in the Phoenix cluster core is consistent
with being isobaric, i.e., $ \sqrt{w} < c_{\rm s}$, where $c_{\rm s} =
1250 (kT/ 5.9 \mbox{ keV})^{1/2}$ km s$^{-1}$ is the adiabatic sound
speed.

\subsection{Imaging simulations}
\label{sec-sim}

To understand the degree of the missing flux of the ALMA SZE data, we
performed imaging simulations using the X-ray pseudo pressure map
(figure \ref{fig-xray}d) of the Phoenix cluster. Given that absolute
values of the pseudo pressure were arbitrary, they were normalized so
that the peak signal corresponds to the Compton $y$-parameter of $y_{\rm
peak}=8\times10^{-4}$, which is a typical value for massive cool core
clusters. To take into account uncertainties of this normalization, we
also examined the cases of $y_{\rm peak}=4 \times 10^{-4}$ and $12
\times 10^{-4}$.  A relativistic correction to the SZE intensity by
\citet{Itoh04} was applied adopting the projected temperature shown in
figure \ref{fig-xray}b at each sky position.

\begin{figure*}[tp]
 \begin{center}
  \includegraphics[height=7.3cm]{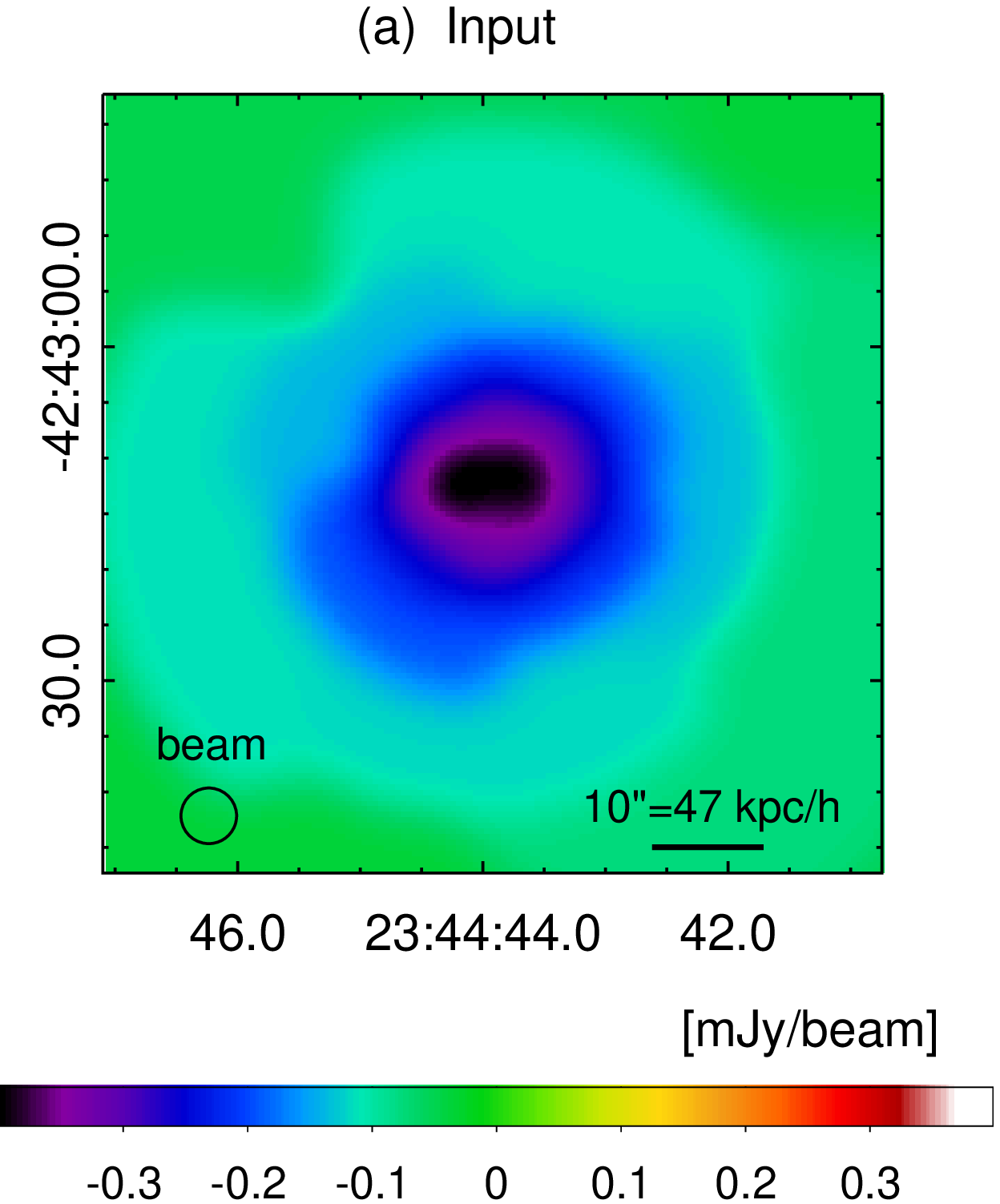}
  \includegraphics[height=7.3cm]{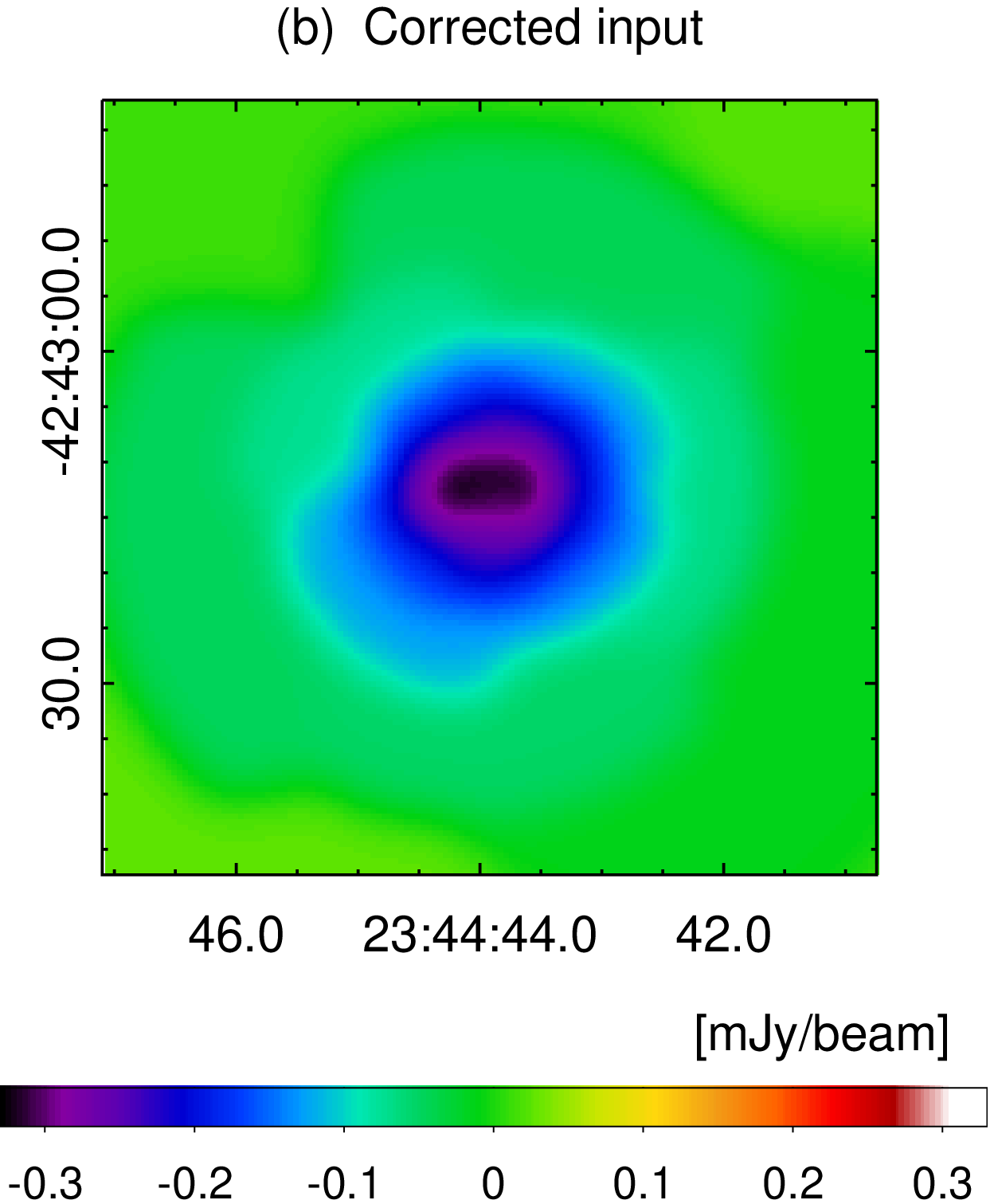}
  \includegraphics[height=7.3cm]{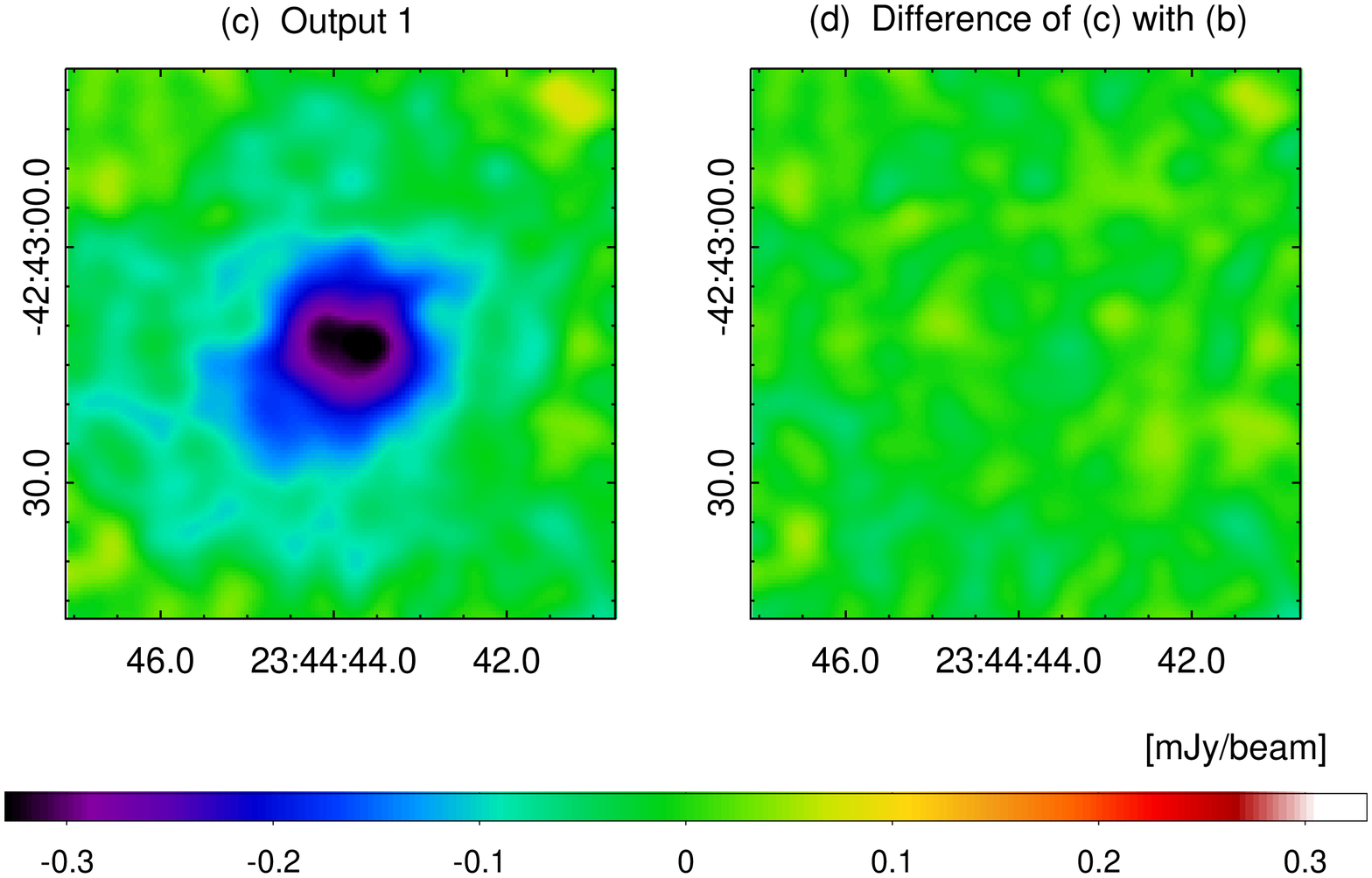}
  \includegraphics[height=7.3cm]{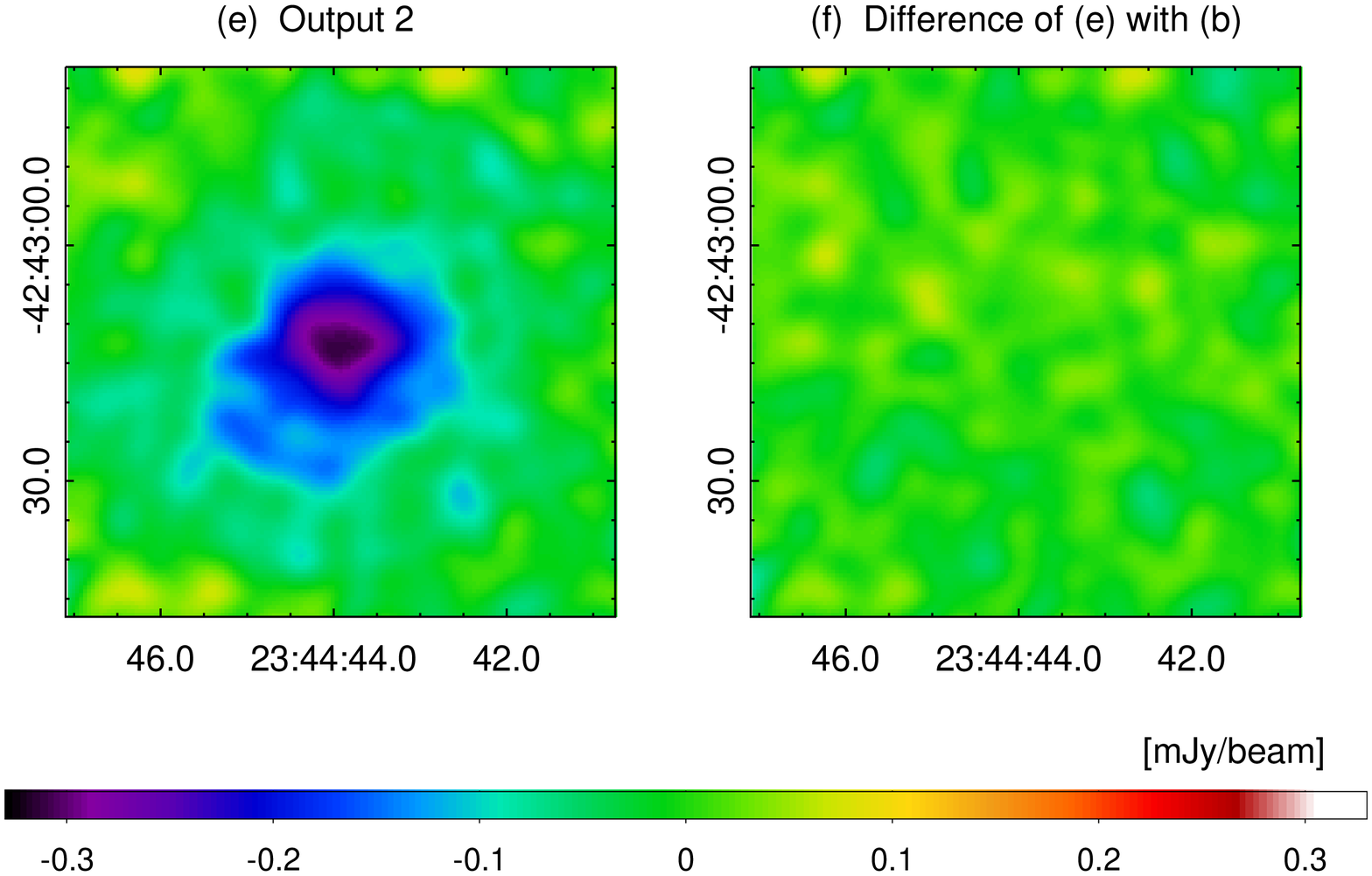}   
 \end{center}
\caption{Mock SZE images of the Phoenix cluster at 92 GHz with
 $y_{\rm peak}=8\times 10^{-4}$. All the images have been smoothed to
 5$''$ FWHM. (a) Input model. (b) Input model to which the correction by
 equation (\ref{eq-conv}) has been applied. (c) Simulation output
 including noise. (d) Difference of panel c with panel b. (e) Same as
 panel c but with different noise realization. (f) Difference of panel e
 with panel b. Note that the range of the color scale is wider in panel
 a than in the other panels.}  \label{fig-sim}
\end{figure*}

We created model images separately at four spectral windows centered at
85, 87, 97, and 99 GHz with an effective bandwidth of 1.875 GHz
each. The pointing directions, the array configuration, the hour angle,
the total effective integration time, and the average precipitable water
vapor were set to match those of each executing block of real
observations. Visibility data were then produced using the CASA task
{\it simobserve} including both instrumental and atmospheric thermal
noise in each spectral window. The rms levels of dirty images are
consistent with the values given in table \ref{tab-obs}. The mock
visibility was deconvolved in the same way as the real data as described
in section \ref{sec-sz}. For each value of $y_{\rm peak}$, we repeated
the above procedure 10 times adopting different noise realizations.
To correct for any potential bias in producing an image at a single
frequency from the data taken over finite bandwidths, the simulation
results are compared with an input model evaluated at the central
frequency 92 GHz.

\begin{figure}
 \begin{center}
  \includegraphics[width=8.3cm]{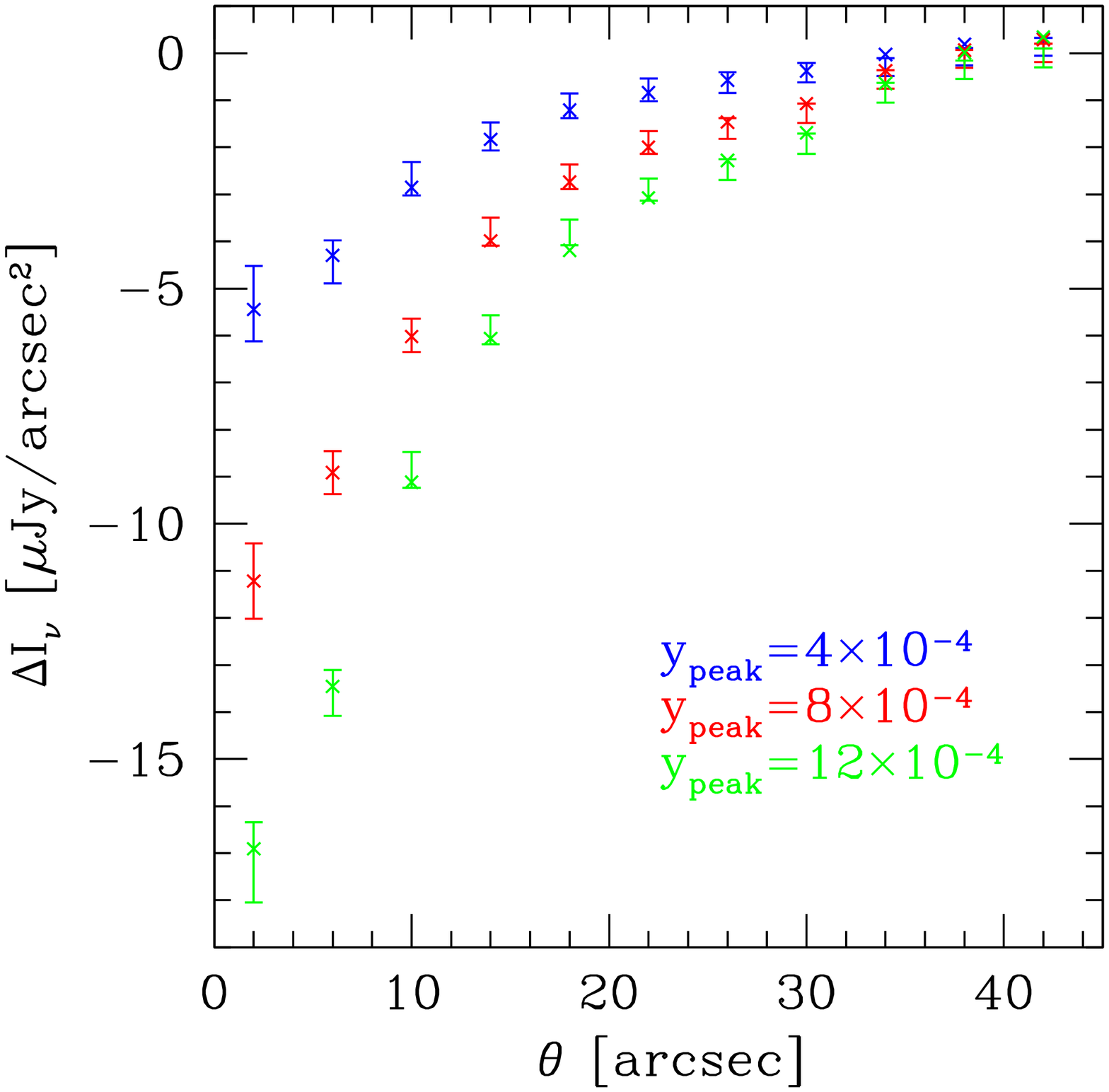} 
 \end{center}
 \caption{Error bars show azimuthally averaged intensity profile of the
 simulation outputs at 92~GHz with $y_{\rm peak}=4\times 10^{-4}$
 (blue), $8 \times 10^{-4}$ (red), and $12\times 10^{-4}$ (green). For
 each value of $y_{\rm peak}$, the results of 10 realizations are
 averaged. Crosses show the same quantity from the input model to which
 the correction by equation (\ref{eq-conv}) has been applied for each
 value of $y_{\rm peak}$.}  \label{fig-prof_sim}
\end{figure}

Figure \ref{fig-sim} compares arbitrarily chosen two realizations of the
simulated images and the input model. The azimuthal average of all
realizations of simulated images are plotted in figure
\ref{fig-prof_sim}.  The simulated images for $y_{\rm peak}=8\times
10^{-4}$ show similar amplitude and spatial extension to the real data
plotted in figures \ref{fig-szmap} and \ref{fig-radprof}. Figures
\ref{fig-sim} and \ref{fig-prof_sim} also illustrate that the simulated
images are in good agreement with the input model once the correction by
equation (\ref{eq-conv}) described below is applied.  The rms values at
$\theta < 45''$ in figures \ref{fig-sim}d and \ref{fig-sim}f are both
0.022 mJy/beam and fully consistent with noise.

\begin{figure}
  \begin{center}
  \includegraphics[width=8.3cm]{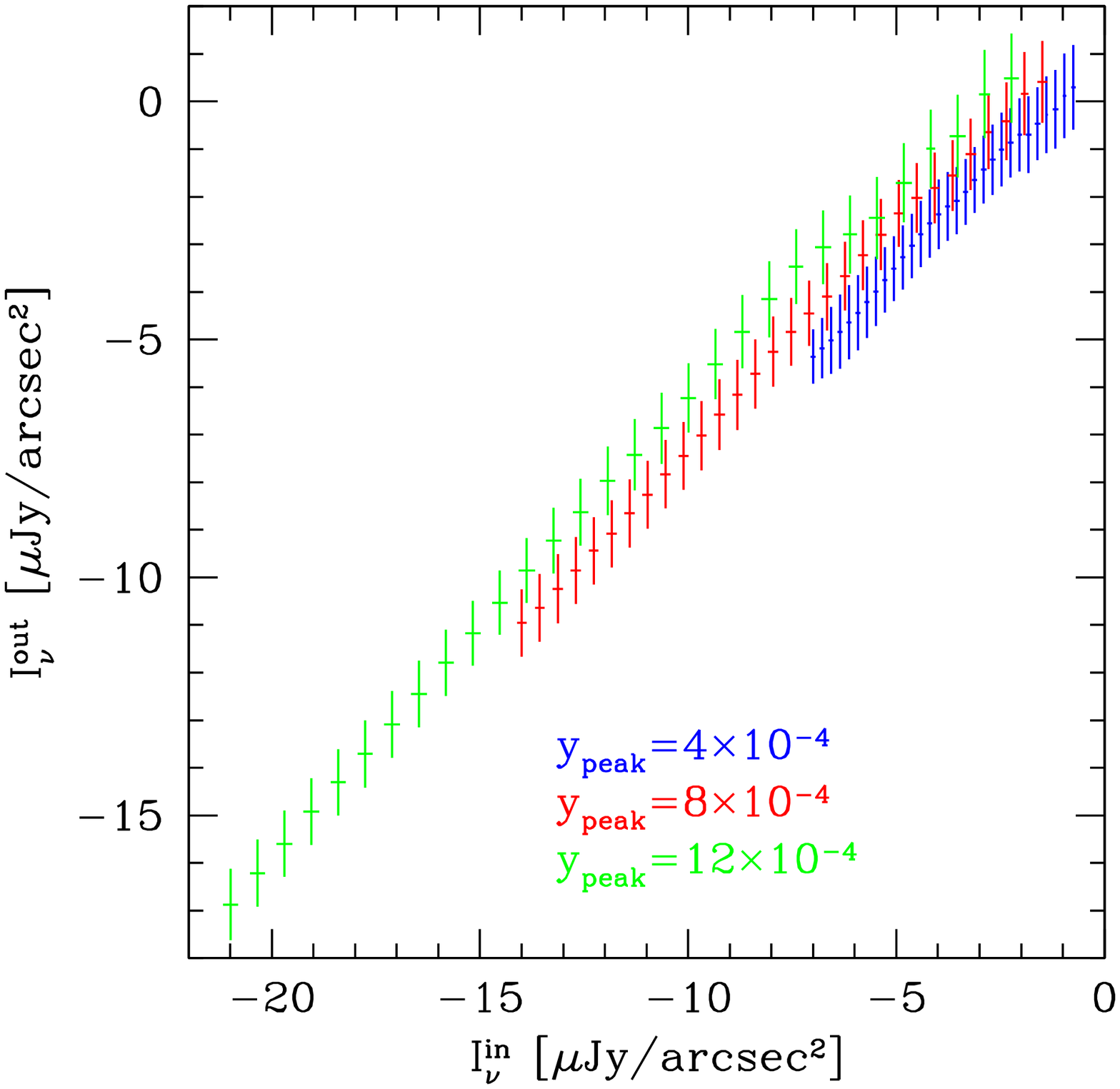}    
  \includegraphics[width=8.3cm]{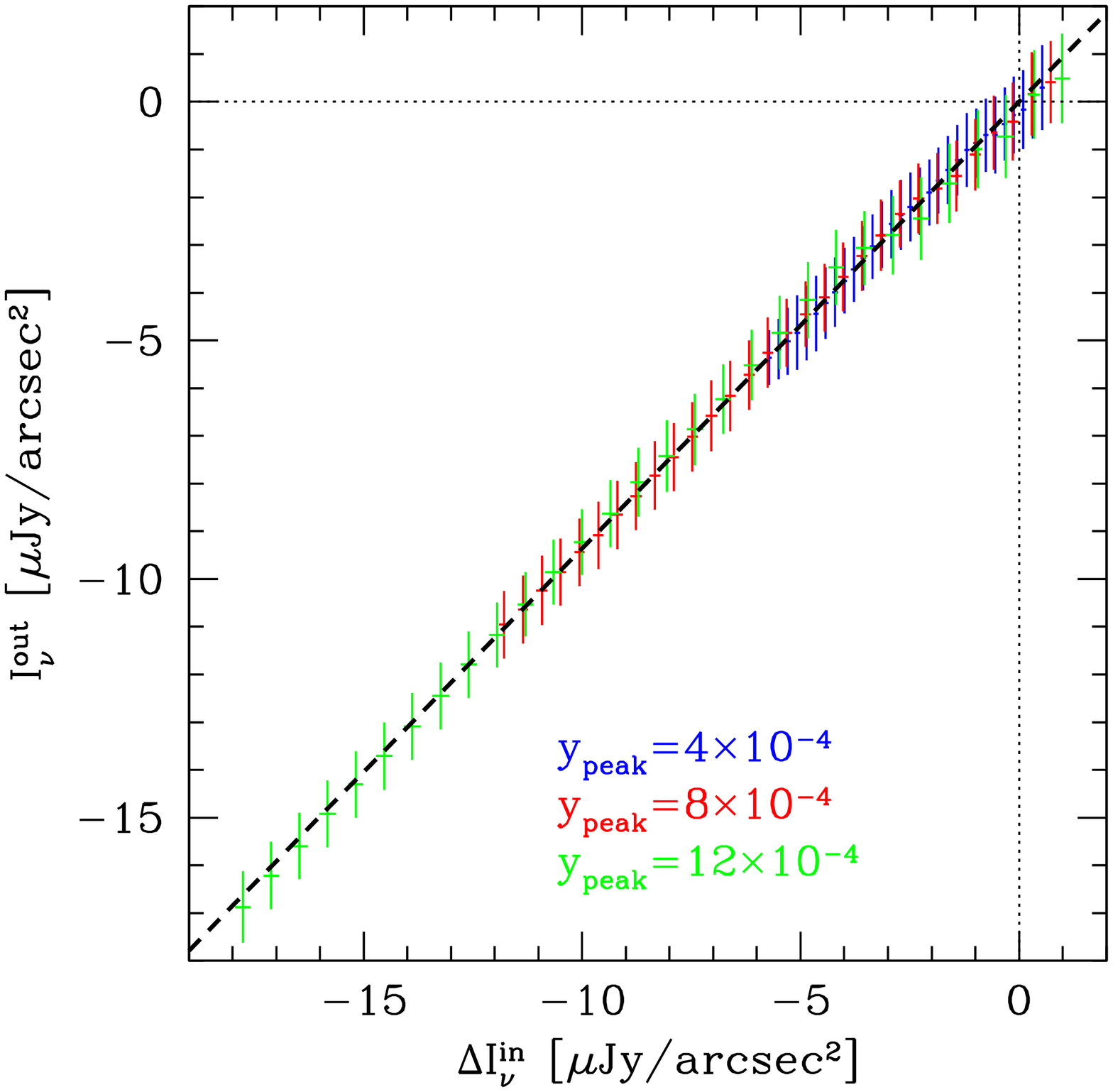} 
  \end{center}
\caption{{\it Top:} Relation between the output intensity and the input
 intensity at 92~GHz from the imaging simulations with $y_{\rm peak}=
 4\times 10^{-4}$ (blue), $8\times 10^{-4}$ (red), and $12\times
 10^{-4}$ (green).  Error bars denote the standard deviation in each
 bin. {\it Bottom:} Same as the left panel, except that the fitted value
 of $c_0/c_1$ in equation (\ref{eq-conv}) has been added to
 $I_{\nu}^{\rm in}$ to give zero intercept in each case. The thick
 dashed line shows the best fitting relation with $c_1=0.94$.}
 \label{fig-conv}
\end{figure}

As in \citet{Kitayama16}, we find that the simulation results on average
follow the linear relation
\begin{eqnarray}
I_{\nu}^{\rm out} = c_1 I_{\nu}^{\rm in} + c_0, 
\label{eq-conv}
\end{eqnarray}
where $I_{\nu}^{\rm out}$ and $I_{\nu}^{\rm in}$ are respectively the
intensities of output and input images at the same sky position. Figure
\ref{fig-conv} shows that the coefficient $c_0$ is sensitive to the
adopted value of $y_{\rm peak}$ whereas $c_1$ remains nearly unchanged
and is close to unity. We hence determined $c_0$ and $c_1$ by fitting
the results of 10 simulation realizations for each value of $y_{\rm
peak}$, where $c_1$ is assumed to be common among different values of
$y_{\rm peak}$. This yielded $c_1=0.94 \pm 0.02$ and $c_0= (1.19 \pm
0.16, ~2.07 \pm 0.20, ~3.02 \pm 0.28)$ $\mu$Jy/arcsec$^2$ for $y_{\rm
peak}=(4\times 10^{-4}, ~8\times 10^{-4}, ~12\times 10^{-4})$,
respectively. The error bars of $I_{\nu}^{\rm out}$ plotted in figure
\ref{fig-conv} show the standard deviation in each bin and are dominated
by statistical errors; each bin contains on average 8000 pixel data
(i.e., 800 per realization). The rms deviation of the mean values of
$(I_{\nu}^{\rm out}, I_{\nu}^{\rm in})$ from the best-fitting relation
is $\Delta I_\nu^{\rm in}=0.17 \mu$Jy/arcsec$^2$ and gives an estimate
of intrinsic deviation from equation (\ref{eq-conv}) apart from the
statistical errors. We regard $\sqrt{2}$ times this value (i.e., $\Delta
I_\nu^{\rm in}=0.24 \mu$Jy/arcsec$^2$) as the 1$\sigma$ systematic
uncertainty in the true intensity when a constant offset (corresponding
to $c_0$) is subtracted. In real observations, the true intensity is
unknown and we will subtract the mean signal at the edge of the emission
profile instead of assuming any specific value of $c_0$ in sections
\ref{sec-y} and \ref{sec-deproj}.

The fact that the simulated images of the Phoenix cluster are well
reproduced by equation (\ref{eq-conv}) with the value of $c_1$ close to
unity (figures \ref{fig-sim}--\ref{fig-conv}) confirms that the observed
ALMA SZE image gives a reasonable representation of differential values
of the true intensity \citep{Kitayama16}. Figure \ref{fig-prof_sim}
further implies that the extended signal is retained out to $\theta \sim
40''$ irrespectively of the adopted value of $y_{\rm peak}$, i.e.,
normalization of the intrinsic signal.

\subsection{Compton $y$-parameter map and the inner pressure profile}
\label{sec-y}

\begin{figure}[t]
  \begin{center}
  \vspace*{-4mm}   
  \includegraphics[width=8.3cm]{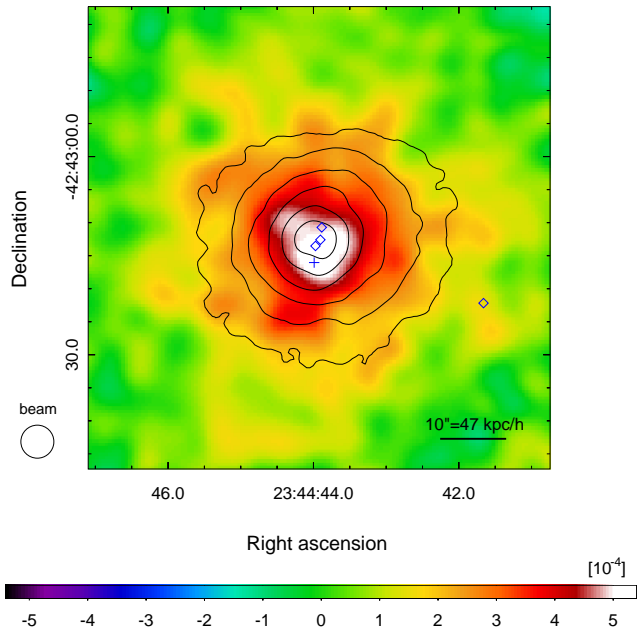} 
  \end{center}
 \caption{Compton $y$-parameter map of the Phoenix cluster at $5''$
 resolution. The map is corrected for the missing flux and the zero
 level is taken at $42''$ from the central AGN. Overlaid are the
 contours of the X-ray surface brightness at 0.7--2.0 keV by Chandra
 corresponding to 64, 32, 16, 8, 4, and 2\% of the peak value, after
 being smoothed by a Gaussian kernel with $2.3''$ FWHM. The positions of
 the SZE peak and subtracted sources are marked by a cross and diamonds,
 respectively.} \label{fig-ymap}
\end{figure}

A high angular resolution SZE image provides a direct probe of the
projected electron pressure, or equivalently, the Compton $y$-parameter,
of the ICM.  Figure \ref{fig-ymap} shows the Compton $y$-parameter map
reconstructed from the observed ALMA image in figure \ref{fig-szmap}
using the results of section \ref{sec-sim}. Specifically, the ALMA image
was divided by the correction factor for the missing flux of 0.94 ($c_1$
in equation [\ref{eq-conv}]) and the mean signal (consistent with zero)
at $\theta = 42''$ is subtracted. At each sky position, the relativistic
correction by \citet{Itoh04} was applied adopting the projected X-ray
spectroscopic temperature shown in figure \ref{fig-xray}b. The plotted values
correspond to the incremental $y$-parameter ($\Delta y$) relative to the
sky positions at $\theta = 42''$. The peak value is $\Delta y = (5.3 \pm
0.4 \mbox{ [statistical]} \pm 0.3 \mbox{ [systematic]})\times 10^{-4}$,
where the systematic error is from the absolute calibration of ALMA
(6\%: \cite{Kitayama16}) and the missing flux correction (0.24
$\mu$Jy/arcsec$^2$: section \ref{sec-sim}). Within the inferred range of
statistical and systematic uncertainties, overall morphology of the
Compton $y$-parameter map does not exhibit deviation from that of
X-rays.

\begin{figure}
 \begin{center}
  \includegraphics[width=8.3cm]{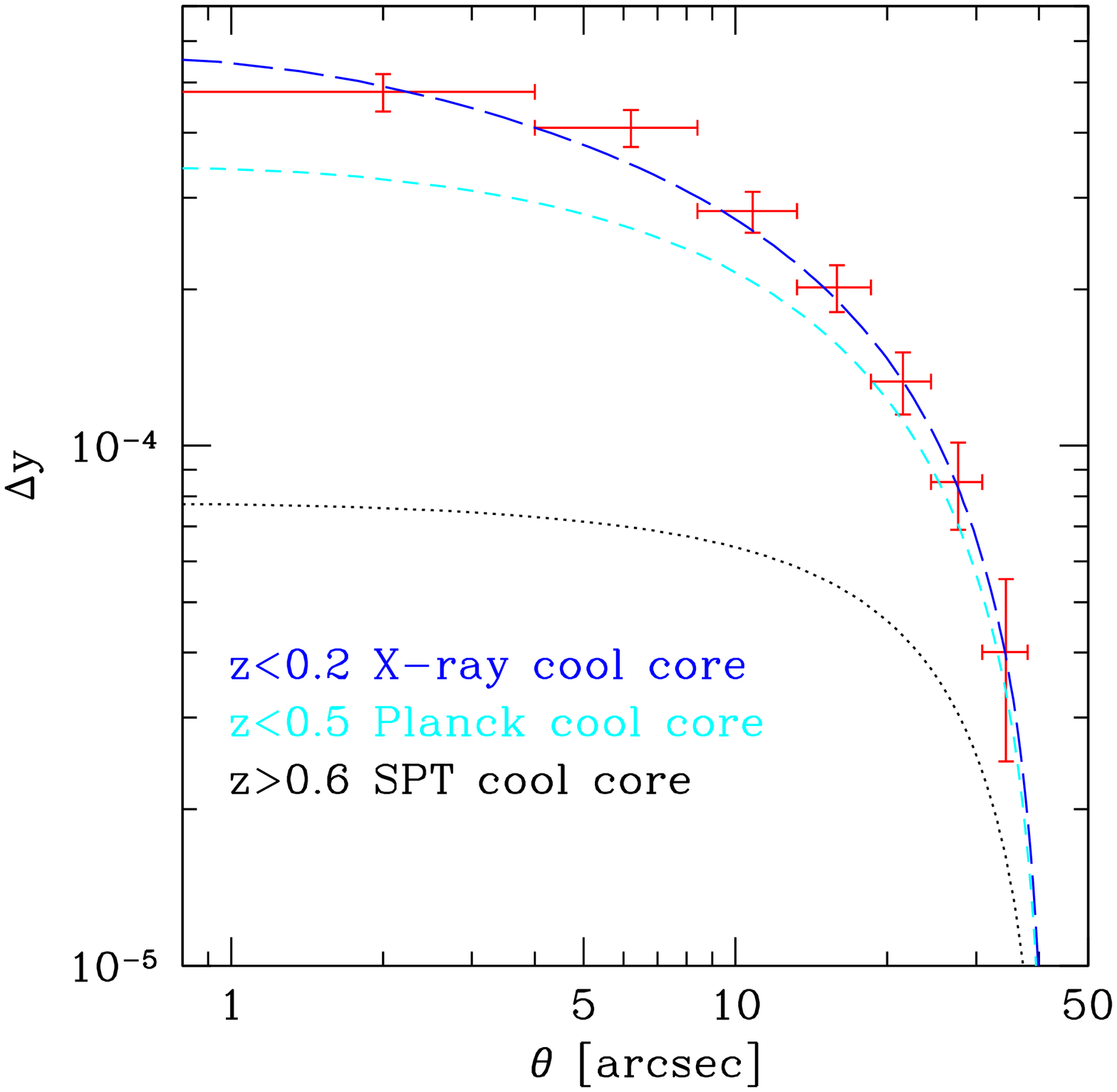} 
 \end{center}
 \caption{Azimuthally averaged Compton $y$-parameter of the Phoenix
cluster.  Overlaid are the expectations from the generalized NFW
pressure profile of X-ray selected cool core clusters at $z<0.2$ (long
dashed line) by \citet{Arnaud10}, Planck selected cool core clusters at
$z < 0.5$ (short dashed line) by \citet{Planck13}, and SPT selected cool
core clusters at $z>0.6$ (dotted line) by \citet{McDonald14b}.  Both the
data points and the expectations are relative to the positions at
$\theta=42''$.} \label{fig-pe}
\end{figure}

Given the regularity of the Compton $y$-parameter map, we compare its
azimuthal average to the mean radial pressure profiles of various galaxy
cluster samples in figure \ref{fig-pe}.  Vertical error
bars include both statistical and systematic errors, whereas horizontal
ones indicate the bin size; the bins are geometrically spaced with the
inner-most bin at $0 < \theta < 4''$ and the size increasing by a factor
of 1.1 so that a statistical error is smaller than $15\%$ in each bin.
The mean radial pressure profiles are computed by integrating along the
line-of-sight the generalized NFW profile \citep{Nagai07} with the
model parameters given in the literature (mentioned below) assuming
$M_{500} = 8.8 \times 10^{14} h^{-1} M_{\odot}$ and $r_{500} = 0.92
h^{-1}$Mpc at $z=0.597$ for the Phoenix cluster\footnote{$M_{500}$ is
the total mass enclosed within the radius $r_{500}$ at which the
enclosed overdensity is 500 times the critical density of the Universe
at that redshift.}  \citep{McDonald12}, being convolved
with the synthesized beam size of ALMA ($2.25'' \times 1.92''$ FWHMs),
and taking a difference with respect to the positions at $42''$ from the
center. We find that the inner pressure profile of the Phoenix cluster
is in good agreement with the expectation from the local cool core
clusters by \citet{Arnaud10} and is clearly steeper than that from more
distant counterparts by \citet{Planck13} and
\citet{McDonald14b}. \citet{McDonald15} showed that the Chandra X-ray
data of this cluster also agree with the pressure profile of the local
cool core clusters of \citet{Arnaud10}. These results further imply that
azimuthally averaged SZE and X-ray data of the Phoenix cluster are
consistent with each other.

\subsection{Deprojected electron temperature and density}
\label{sec-deproj}

\begin{figure*}[tp]
 \begin{center}
  \includegraphics[width=8.3cm]{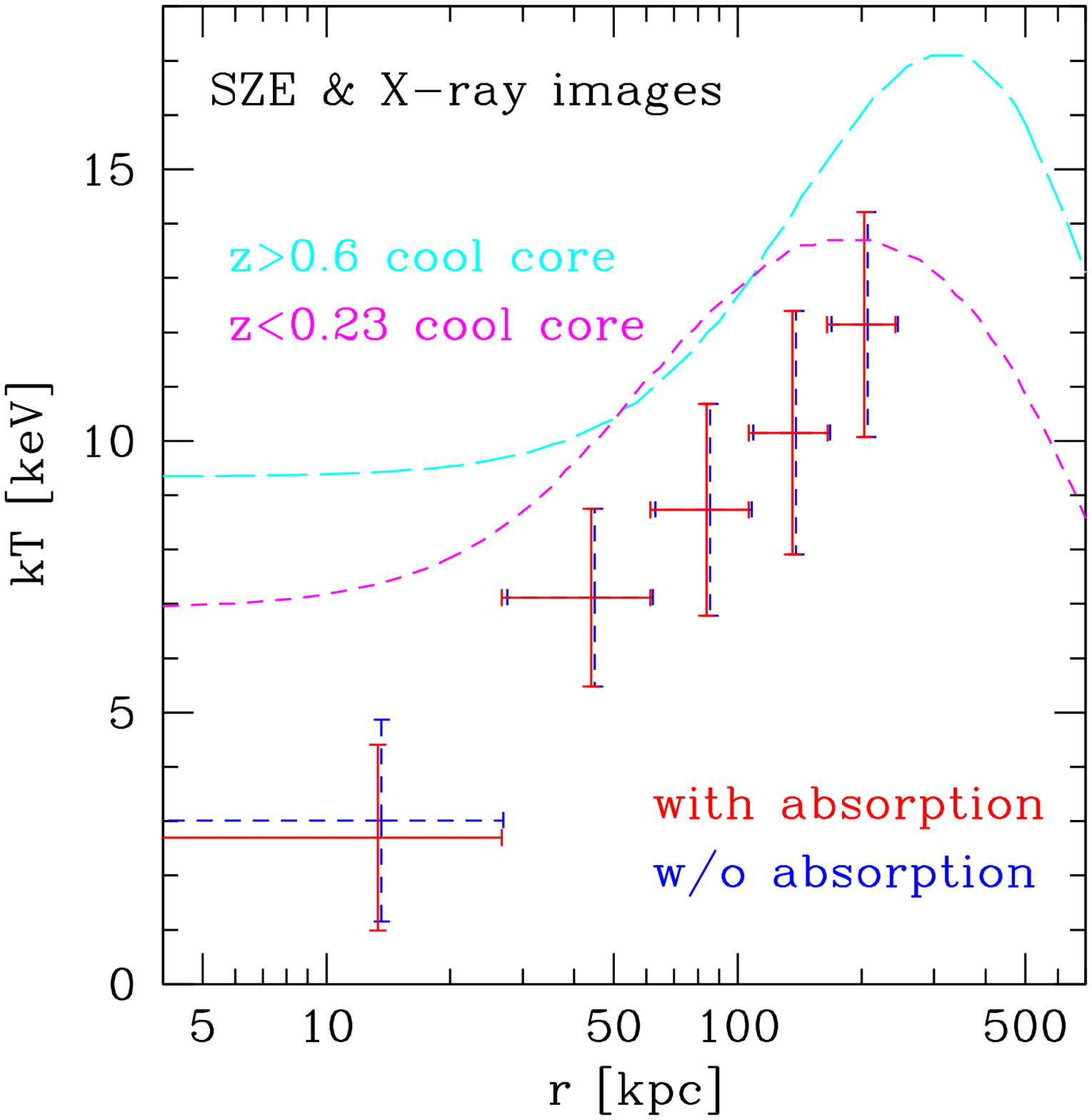}
  \includegraphics[width=8.3cm]{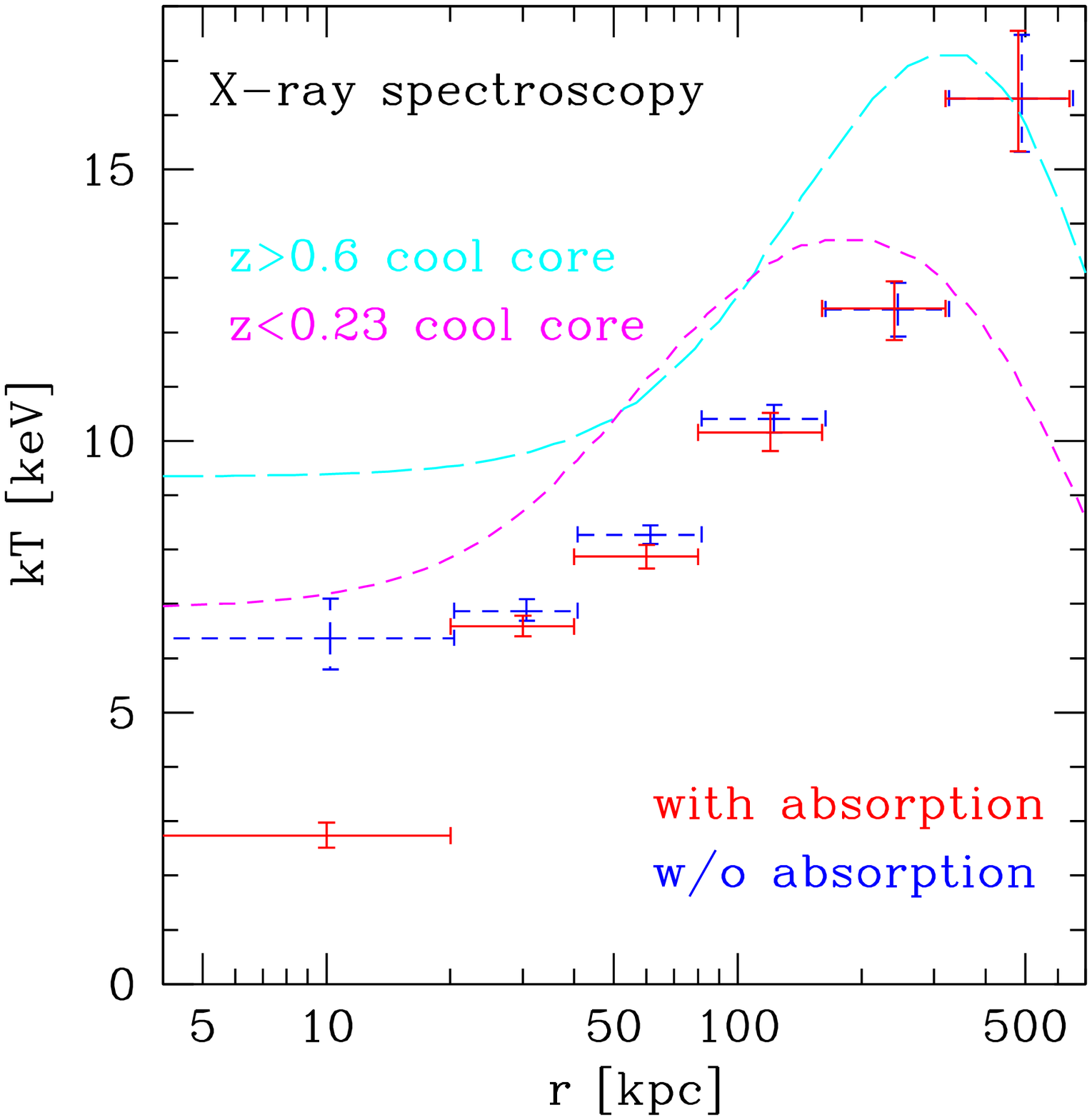}  
 \end{center}
 \caption{Deprojected temperature profiles of the
 Phoenix cluster from SZE and X-ray images (left panel) and from X-ray
 spectroscopy (right panel) assuming $h=0.7$.  The error bars show the
 results with (solid) or without (dashed) intrinsic absorption inside
 the cluster.  For clarity, dashed error bars are slightly shifted
 horizontally.  Lines indicate the mean profiles of cool core clusters
 at $z>0.6$ (long-dashed) by \citet{McDonald14b} and at $z<0.23$
 (short-dashed) by \citet{Vikhlinin06}.  } \label{fig-tdeproj}
\end{figure*}

A combination of SZE and X-ray images provides a measure of temperature
and density of the ICM independently of the X-ray spectroscopy (e.g.,
\cite{Kitayama04, Nord09, Basu10}).  Assuming spherical symmetry and
$h=0.7$ for this particular purpose, we performed non-parametric
deprojection of electron density $n_{\rm e}$ and temperature $T$ as
described in detail below. A major advantage of this method is that the
impact of X-ray absorbing gas described in section \ref{sec-xray} is
minimized. If the absorbed and unabsorbed X-ray brightness is related by
$I_{\rm X}^{\rm abs} = \alpha I_{\rm X}^{\rm unabs}$ ($0 \leq \alpha
\leq 1$), deprojected density and temperature vary approximately as
$n_{\rm e}^{\rm abs} = n_{\rm e}^{\rm unabs}/\sqrt{\alpha}$ and $T^{\rm
abs} = T^{\rm unabs} \sqrt{\alpha}$, respectively. The intrinsic
absorption inferred in section \ref{sec-xray} gives $\alpha \sim 0.75$
at $\theta < 1.5''$ in the 0.7--2.0 keV band, implying that the central
density and temperature are expected to change only by $ \sim 13\%$.

We first took geometrically-spaced bins on the sky with the inner-most
bin at $0 < \theta < 4''$ (corresponding to the deprojected spherical
shell of $0 < r < 26.7$ kpc) and the size increasing by
a factor of 1.3 so that a statistical error is smaller than $10\%$ in
each bin. Systematic errors from the flux calibration of ALMA (6\%),
the missing flux correction of the SZE (0.24 $\mu$Jy/arcsec$^2$), and
the effective area of Chandra ACIS-I (4\%) were added in quadrature to
the statistical error in each bin. We then fitted the
volume averaged brightness in each bin of the SZE at $\theta < 40''$
and of the 0.7--2.0 keV X-rays at $\theta < 100''$ together varying the
temperature and the density in each spherical shell. As the temperature
at $r > 250$ kpc ($\theta > 37''$) cannot be constrained by the ALMA
data, it was fixed at the projected mean value of 15.8 keV by the X-ray
spectral analysis described in section \ref{sec-xray}; we checked that
the deprojected quantities at $r < 100$ kpc are essentially unchanged by
this assumption. For the SZE, we modeled and fitted the incremental
brightness relative to the bin centered at $\theta=43.6''$; 
the missing flux correction factor of 0.94 and the temperature-dependent
relativistic correction were applied to the model brightness in
each spherical shell.  The X-ray emissivity was computed by SPEX
v3.0.5.00 \citep{Kaastra96}, fixing the metal abundance at 0.35 times
the Solar value. To examine the impact of intrinsic absorption within
the Phoenix cluster, the hydrogen column density in the inner-most bin
was assumed to be either $N_{\rm H, int} =6.7 \times 10^{21}$ cm$^{-2}$
or $N_{\rm H, int} =0$; the former value was chosen as a limiting case
of strong absorption throughout the inner 27 kpc.  Note that the X-ray
emission from the central AGN is negligible in the 0.7--2.0 keV band
used in the analysis.

For comparison, we separately performed a deprojection analysis using
solely the X-ray data. The 0.7--7.0 keV spectra in six circular annuli
at $\theta=0''-3''$, $3''-6''$, $6''-12''$, $12''-24''$, $24''-48''$,
and $48''-96''$ were fitted together using the model {\it projct}
implemented in XSPEC version 12.10.0e \citep{Arnaud96}. For simplicity,
only statistical errors were taken into account in this particular
analysis. The position-dependent free parameters were the same as those
described in section \ref{sec-xray} except that they were assigned to
each of six spherical shells corresponding to the above mentioned
annuli.  The spectral index, column density of an obscuring torus, and
the flux ratio between the 6.4 keV line and the power-law continuum of
the central AGN were fixed at their best-fit values obtained in section
\ref{sec-xray}.

\begin{figure*}[tp]
 \begin{center}
  \includegraphics[width=8.3cm]{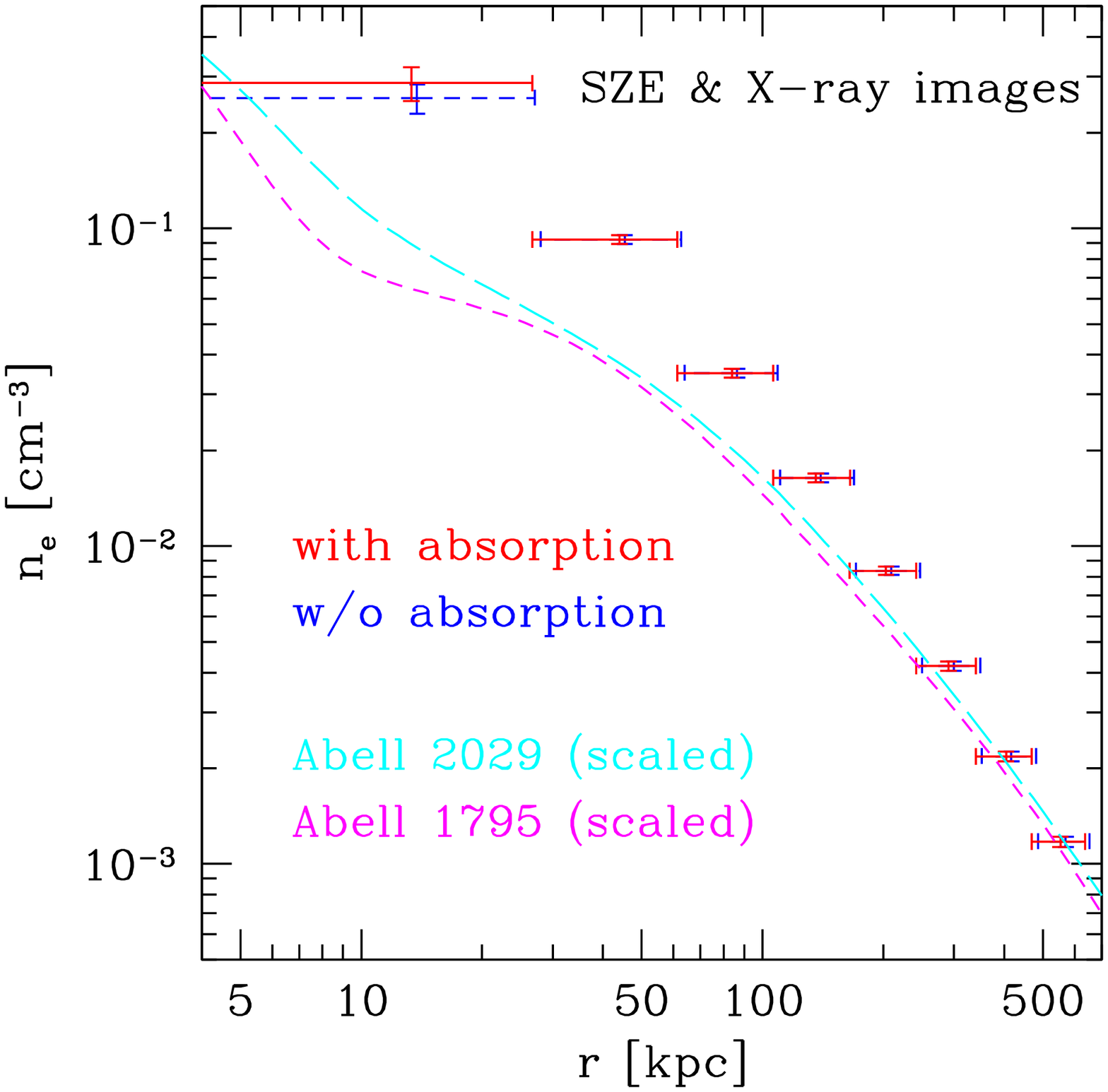}
  \includegraphics[width=8.3cm]{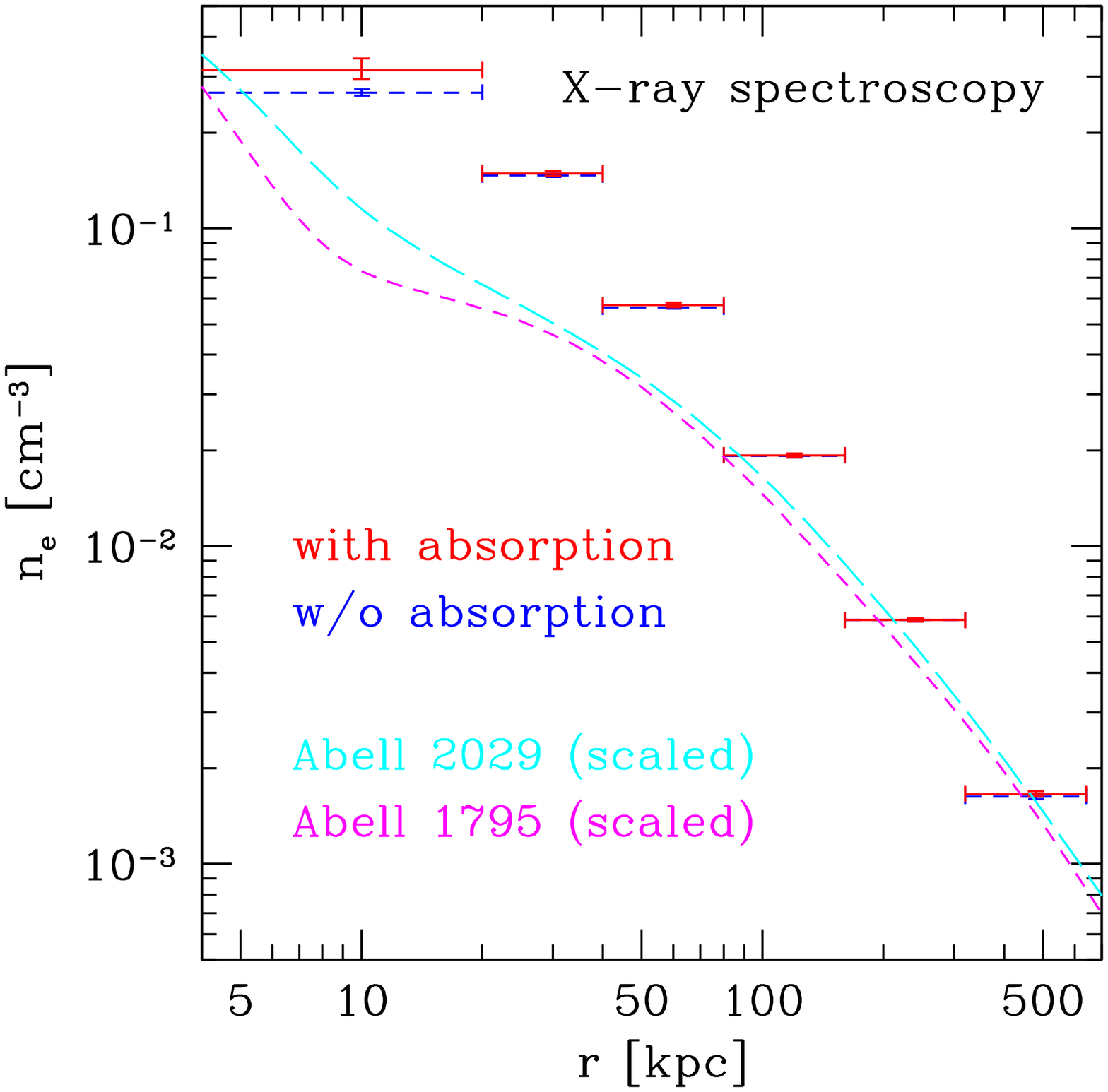}  
 \end{center}
 \caption{Same as figure \ref{fig-tdeproj} except for showing
 deprojected electron density.  Lines indicate the profiles of local
 cool core clusters Abell 2029 (long-dashed) and Abell 1795
 (short-dashed) by \citet{Vikhlinin06}; both profiles are scaled to
 match the radius and redshift of the Phoenix cluster as described in
 the text.  In the left panel, the data points at $r>250$ kpc are
 obtained solely from the X-ray image assuming $kT=15.8$ keV.} 
 \label{fig-nedeproj}
\end{figure*}

Figure \ref{fig-tdeproj} shows deprojected temperatures
obtained by two different methods. As expected, the results from the SZE
and X-ray images are insensitive to intrinsic absorption;
the temperature at $r<27$ kpc is $2.70 \pm 1.71$ keV and
$3.01 \pm 1.86$ keV with and without intrinsic absorption,
respectively. Those solely from the X-ray data are much more sensitive,
albeit with smaller statistical errors; the temperature at $r<20$ kpc
changes by more than a factor of two from $2.73^{+0.24}_{-0.22}$ keV to
$6.37^{+0.73}_{-0.58}$ keV once intrinsic absorption is neglected. If
intrinsic absorption is taken into account, the temperature profile from
the latter deprojection method comes to a good agreement with that from
the former.

Also plotted for reference in figure \ref{fig-tdeproj} are the mean
profiles of local and distant cool core clusters
\citep{Vikhlinin06,McDonald14b}, using the average temperature within
$r_{500}$ of this cluster, $kT_{500}=12.6$ keV, expected from the X-ray
scaling relation by \citet{Reichert11}. We find that the deprojected
temperatures at $r<100$ kpc are systematically lower than the average
profiles of both local and distant clusters. The inferred central
temperature is $\sim 5$ times lower than the mean temperature ($\sim 16$
keV) at $r>300$ kpc. This is a much larger factor of reduction than
observed in other clusters and supports that radiative cooling is
efficient in the Phoenix cluster.

Given that the pressure profile of the Phoenix cluster well matches that
of local cool core clusters (section \ref{sec-y}), gas density should
have been enhanced to compensate for the reduced temperature. Figure
\ref{fig-nedeproj} shows that this is indeed the case; overlaid are the
best-fit density profiles by \citet{Vikhlinin06} of typical local cool
core clusters, Abell 2029 and Abell 1795. To correct for different size
and redshift of individual clusters, these profiles have been scaled so
that $r$ gives the same fraction of $r_{500}$ as the Phoenix cluster and
that the normalization of $n_{\rm e}$ evolves to $z=0.597$ by the same
fraction as the critical density of the Universe.  Gas density of the
Phoenix cluster is in good agreement with those of Abell 2029 and Abell
1795 at $r> 0.2 r_{500} \sim 300$ kpc, whereas it is systematically
higher at smaller radii. Figure \ref{fig-nedeproj} also
confirms that density profiles obtained from two deprojection methods
are consistent with each other.

\begin{figure*}[tp]
 \begin{center}
  \includegraphics[width=8.3cm]{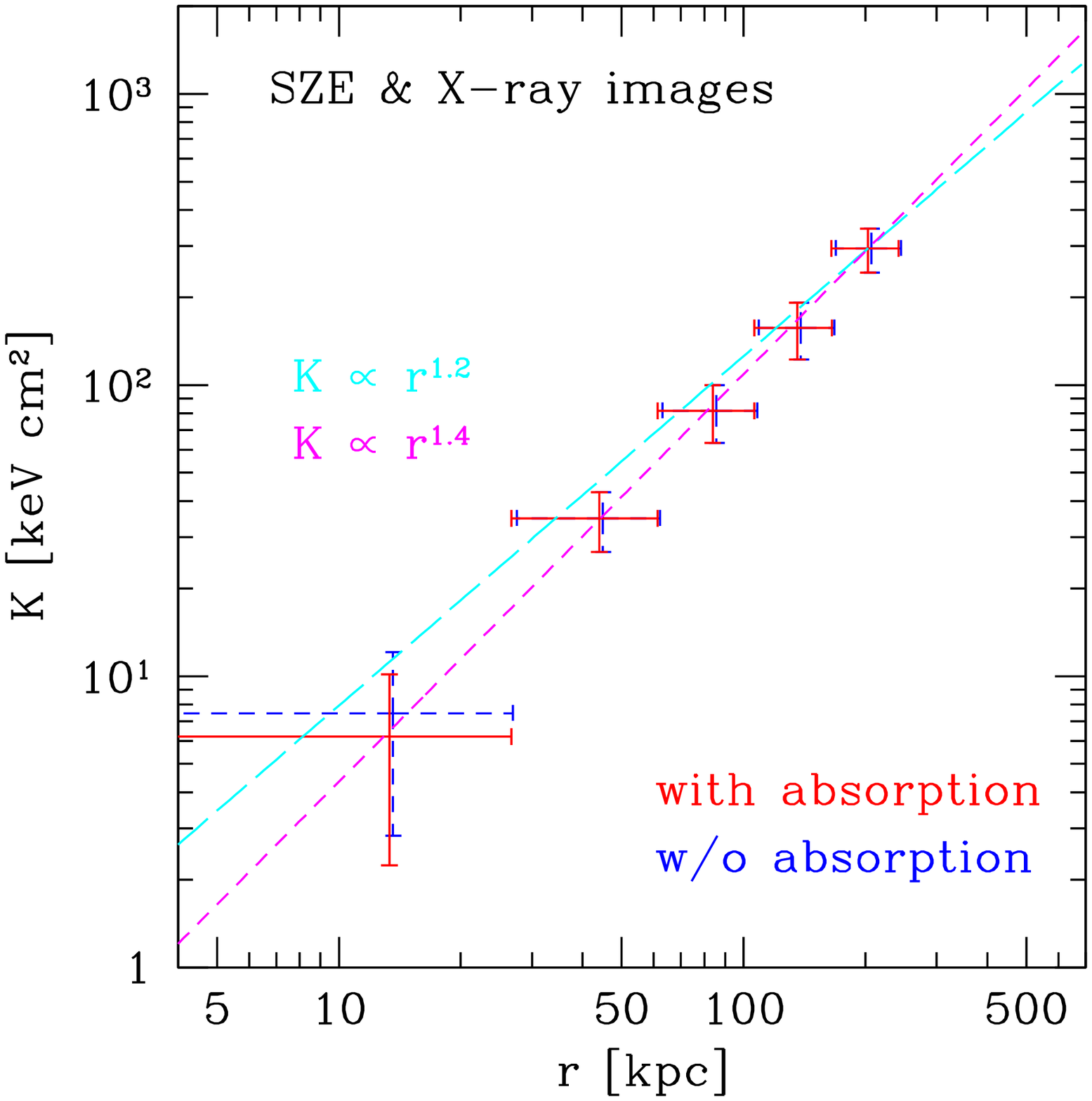}
  \includegraphics[width=8.3cm]{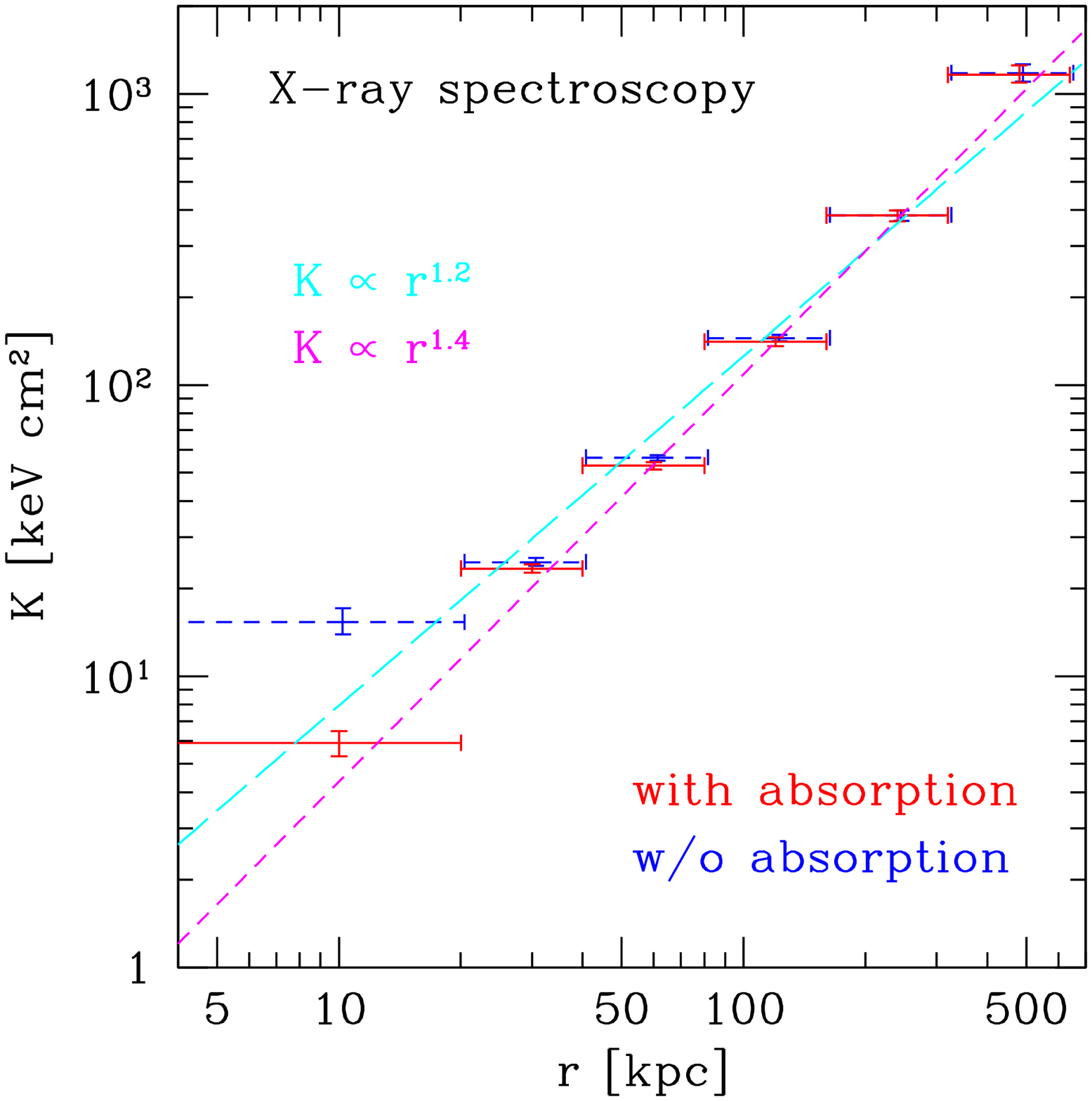}  
 \end{center}
 \caption{Same as figure \ref{fig-tdeproj} except for showing
 entropy. Lines indicate the model profiles $K\propto r^{1.2}$
 (long-dashed) \citep{Voit05} and $K\propto r^{1.4}$ (short-dashed)
 \citep{Voit11}, both normalized to match the data point in the
 outer-most bin in the left panel.}  \label{fig-kdeproj}
\end{figure*}

Figure \ref{fig-kdeproj} further illustrates that the entropy $K\equiv
kT n_{\rm e}^{-2/3}$ decreases to $<10$ keV cm$^2$ at $r<27$ kpc.  Also
plotted are a model $K\propto r^{1.2}$ which tends to give a lower limit
to non-radiative clusters at $r \ltsim 0.5 r_{500}$ \citep{Voit05} and a
prediction $K\propto r^{1.4}$ for the steady-state cooling flow
\citep{Voit11}; both profiles are normalized to match the data point in
the outer-most bin in the left panel. The observed entropy profile 
shows a better agreement with $K\propto r^{1.4}$ than $K\propto r^{1.2}$
and strongly supports that radiative cooling is efficient in the
Phoenix cluster.

\begin{figure*}[tp]
 \begin{center}
  \includegraphics[width=8.3cm]{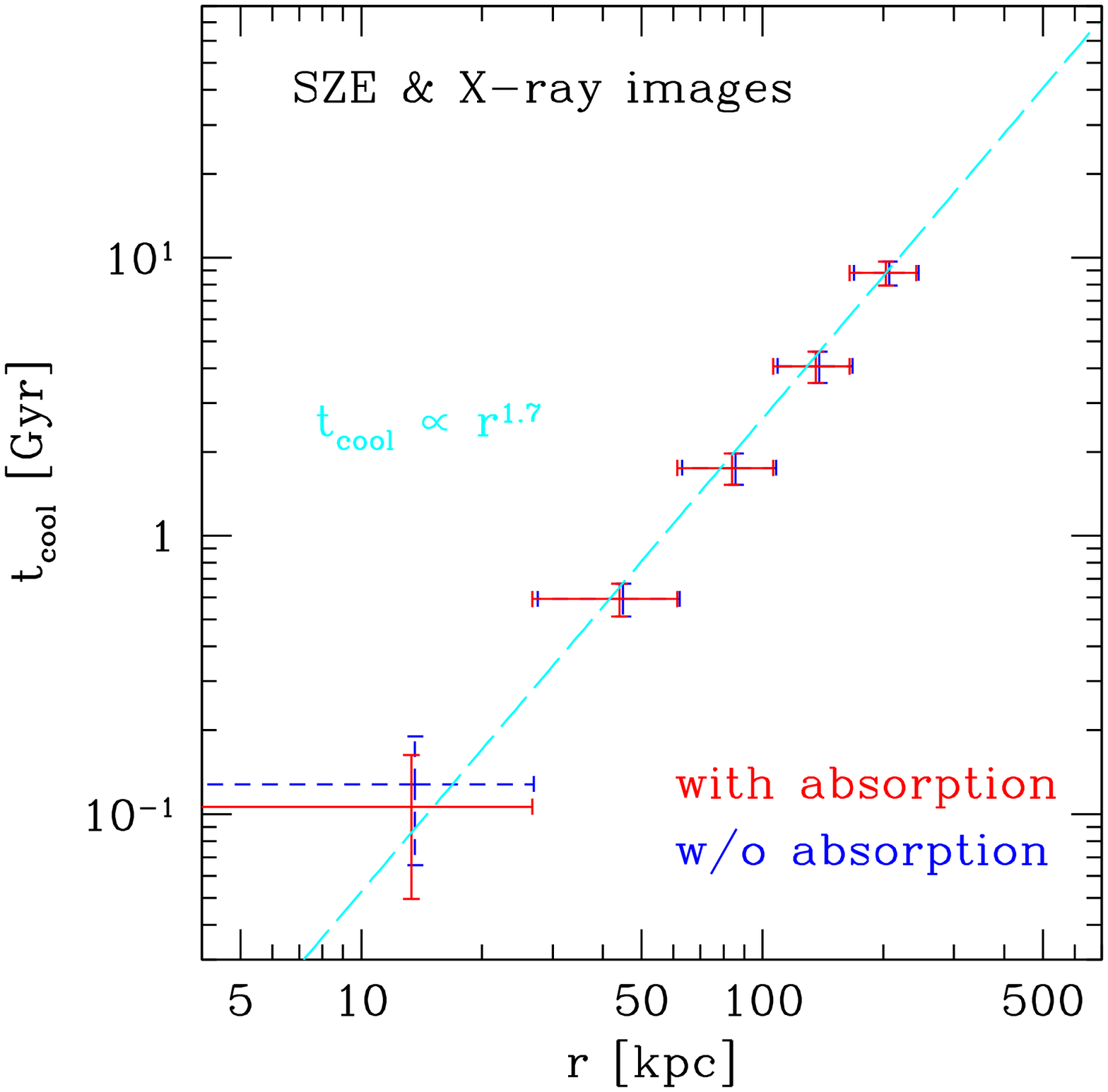}
  \includegraphics[width=8.3cm]{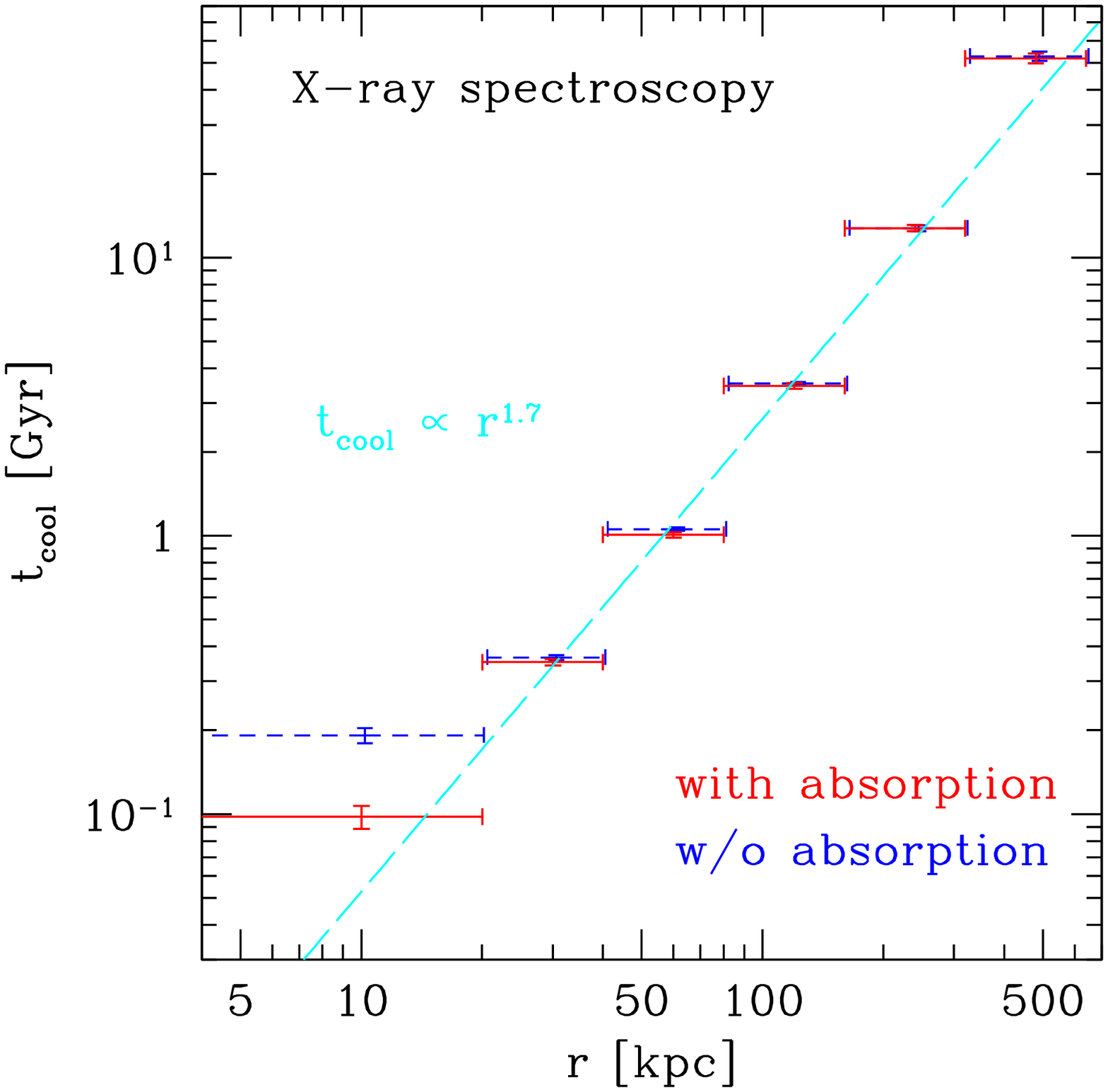}   
 \end{center}
 \caption{Same as figure \ref{fig-tdeproj} except for
 showing the radiative cooling time. The dashed line indicates the
 relation $t_{\rm cool} \propto r^{1.7}$, normalized to match the data
 point in the outer-most bin in the left panel.} \label{fig-tcdeproj}
\end{figure*}

Finally, we plot in Figure \ref{fig-tcdeproj} the radiative cooling
time, i.e., total thermal energy of the gas divided by the bolometric
luminosity in each radial bin. As the bolometric luminosity of the gas
at $kT \gtsim 1$ keV is dominated by X-rays, it is computed by
integrating the rest-frame X-ray emissivity provided by SPEX over photon
energies (e.g., \cite{Schure09}). The radial profile of the cooling time
is well represented by a single power-law of $t_{\rm cool} \propto
r^{1.7}$ and drops to $\sim 0.1$ Gyr at $r<27$ kpc.

\subsection{How much gas is cooling in the Phoenix cluster?}

The results in section \ref{sec-deproj} immediately yield the amount of
$\sim 3$ keV gas within $r=27$ kpc to be $M_{\rm
cool}=(6.4\pm 0.9) \times 10^{11} M_\odot$. This corresponds to $\sim
20\%$ of the stellar mass $\sim 3 \times 10^{12} M_\odot$ in the central
galaxy of the Phoenix cluster \citep{McDonald12, McDonald13}. If the
mass deposition rate at $kT \gtsim 3 $ keV is $\dot{M}_{\rm cool} \sim
2000 ~M_\odot$ yr$^{-1}$ as indicated by the X-ray data (e.g.,
\cite{McDonald13,Ueda13}), the cool gas should have accumulated over the
period
\begin{eqnarray}
\tau_{\rm acc} = \frac{M_{\rm cool}}{\dot{M}_{\rm
 cool} } = (0.32 \pm 0.05)  \left(\frac{\dot{M}_{\rm
			     cool} }{2000 ~M_\odot
			     \mbox{yr}^{-1}}\right)^{-1} \mbox{Gyr}. 
 \label{eq-tau}
\end{eqnarray}
This is feasible because $t_{\rm cool} < \tau_{\rm acc}$ is achieved out
to $r > 27$ kpc (figure \ref{fig-tcdeproj}). The required inflow
velocity of the gas
\begin{eqnarray}
 v =\frac{\dot{M}_{\rm cool}}{4\pi r^2 \rho } &=& 62 
  \left(\frac{r}{30 \mbox{~kpc}}\right)^{-2}
  \left(\frac{n_{\rm e}}{10^{-1} \mbox{cm}^{-3}}\right)^{-1}
    \nonumber \\ && \times 
  \left(\frac{\dot{M}_{\rm
			     cool} }{2000 ~M_\odot
			     \mbox{yr}^{-1}}\right)
  \mbox{~km s}^{-1}
\end{eqnarray}
is also much smaller than the adiabatic sound speed $c_s \simeq 900 (kT/
3 \mbox{ keV})^{1/2}$ km s$^{-1}$ over the range of radii considered in
this paper. 

Our results consistently imply that radiative cooling is hardly
suppressed down to $kT \sim 3$ keV. At even lower temperatures, previous
works infer that cooling may weaken moderately with the mass deposition
rate of $\dot{M}_{\rm cool} = 130-480~M_\odot$ yr$^{-1}$ for the gas
below 2 keV \citep{Pinto18} and the star formation rate of $\dot{M}_{\rm
SF} = 400 - 900 ~M_\odot$ yr$^{-1}$ in the central galaxy
\citep{McDonald12, McDonald13, Mittal17}. If this is the case, the time
required for the above mentioned cool gas to turn into stars will be
\begin{eqnarray}
\tau_{\rm SF} = \frac{M_{\rm cool}}{\dot{M}_{\rm
 SF} } = (1.3 \pm 0.2)  \left(\frac{\dot{M}_{\rm
			     SF} }{500 ~M_\odot
			     \mbox{yr}^{-1}}\right)^{-1} 
 \mbox{Gyr}, 
 \label{eq-tsf}
\end{eqnarray}
i.e., the stellar mass of the central galaxy is expected to increase by
$\sim 20 \%$ over this period.

\section{Conclusions}

We have presented the SZE image of the Phoenix galaxy cluster at
$z=0.597$ taken by ALMA in Band 3. The SZE is imaged at $5''$ resolution
(or 23 $h^{-1}$kpc) within 200 $h^{-1}$kpc from the central AGN with the
peak S/N exceeding 11. Combined with the Chandra X-ray image, the ALMA
SZE data further allow for non-parametric deprojection of electron
temperature, density, and entropy. Our method can minimize
contamination by the central AGN and the X-ray absorbing gas within the
cluster, both of which largely affect the X-ray spectrum.

We find no significant asymmetry or disturbance in the SZE image within
the current measurement errors.  The detected signal shows much higher
central concentration than other distant clusters and agrees well with
the average pressure profile of local cool-core
clusters. Unlike typical clusters at any redshift,
gas temperature drops by at least a factor of 5 toward the center. In
the inner $20 ~h^{-1}$ kpc, we identify the presence of $\sim 6 \times
10^{11} M_\odot$ cool gas with $kT \sim 3 $ keV, the
amount of which corresponds to $\sim 20\%$ of the stellar mass in the
central galaxy. The low entropy ($\sim 10 $ keV cm$^2$) and the short
cooling time ($\sim 0.1 $ Gyr) of this gas further corroborates that
radiative cooling is hardly suppressed between $kT
\sim 16 $ keV and $kT \sim 3 $ keV. Taken together, our results imply
that the gas is cooling efficiently and nearly isobarically down to the
inner $20 h^{-1}$ kpc in the Phoenix cluster.

\hfill 

\begin{ack}

We thank the anonymous referee and Luca Di Mascolo for their helpful
comments.  This paper makes
use of the following ALMA data: ADS/JAO.ALMA\#2015.1.00894.S.  The
scientific results of this paper are based in part on data obtained from
the Chandra Data Archive: ObsID 13401, 16135, 16545, 19581, 19582,
19583, 20630, 20631, 20634, 20635, 20636, and 2079.  ALMA is a
partnership of ESO (representing its member states), NSF (USA) and NINS
(Japan), together with NRC (Canada), MOST and ASIAA (Taiwan), and KASI
(Republic of Korea), in cooperation with the Republic of Chile. The
Joint ALMA Observatory is operated by ESO, AUI/NRAO and NAOJ. The
National Radio Astronomy Observatory is a facility of the National
Science Foundation operated under cooperative agreement by Associated
Universities, Inc. This work was supported by Japan Society for the
Promotion of Science (JSPS) KAKENHI Grant Numbers JP15H03639 (TA),
JP15H05892 (MO), JP15K17614 (TA), JP17H01110 (TA), JP17H06130 (KK, RK),
JP18K03693 (MO), and JP18K03704 (TK), by the NAOJ ALMA Scientific
Research Grant Number 2017-06B (KK), by the Ministry of Science and
Technology of Taiwan (grant MOST 106-2628-M-001-003-MY3), and by
Academia Sinica (grant AS-IA-107-M01).

\end{ack}

\end{document}